\documentstyle[preprint,eqsecnum,aps]{revtex}
\begin{document}


\def\Rnum{{\bf R}}
\def\Cnum{{\bf C}}

\def\eqref#1{(\ref{#1})}
\def\eqrefs#1#2{(\ref{#1}) and~(\ref{#2})}
\def\eqsref#1#2{(\ref{#1}) to~(\ref{#2})}

\def\Eqref#1{Eq.~(\ref{#1})}
\def\Eqrefs#1#2{Eqs.~(\ref{#1}) and~(\ref{#2})}
\def\Eqsref#1#2{Eqs.~(\ref{#1}) to~(\ref{#2})}

\def\secref#1{Sec.~\ref{#1}}
\def\secrefs#1#2{Secs.~\ref{#1} and~\ref{#2}}
\def\secsref#1#2{Secs.~\ref{#1} to~\ref{#2}}

\def\appref#1{App.~\ref{#1}}

\def\Ref#1{Ref.\cite{#1}}
\def\Refs#1{Refs.\cite{#1}}

\def\Cite#1{${\mathstrut}^{\cite{#1}}$}

\def\tableref#1{Table~\ref{#1}}

\def\figref#1{Fig.~\ref{#1}}

\hyphenation{Eq Eqs Sec App Ref Fig}

\def\demo #1. #2\par{\medbreak\noindent{\bf#1.\enspace}{\rm#2}\par\medbreak}

\def\EQ{\begin{equation}}
\def\EQs{\begin{eqnarray}}
\def\endEQ{\end{equation}}
\def\endEQs{\end{eqnarray}}

\def\eqtext#1{\hbox{\rm{#1}}}

\def\proclaim#1{\medbreak
\noindent{\it {#1}}\par\medbreak}
\def\Proclaim#1#2{\medbreak
\noindent{\bf {#1}}{\it {#2}}\par\medbreak}


\def\fewquad{\qquad\qquad}
\def\severalquad{\qquad\fewquad}
\def\manyquad{\qquad\severalquad}
\def\manymanyquad{\manyquad\manyquad}

\def\sp#1{\vskip #1pt}

\def\sub#1{
\setbox1=\hbox{{$\scriptscriptstyle #1$}} 
\dimen1=0.6\ht1
\mkern-2mu \lower\dimen1\box1 \hbox to\dimen1{\box1\hfill} }

\def\eqtext#1{\hbox{\rm{#1}}}

\def\endproof{
\setbox2=\hbox{{$\sqcup$}} \setbox1=\hbox{{$\sqcap$}} 
\dimen1=\wd1
\box2\kern-\dimen1 \hbox to\dimen1{\box1} }

\def\mstrut{\mathstrut}
\def\hp#1{\hphantom{#1}}

\def\mixedindices#1#2{{\mstrut}^{\mstrut #1}_{\mstrut #2}}
\def\downindex#1{{\mstrut}^{\mstrut}_{\mstrut #1}}
\def\upindex#1{{\mstrut}_{\mstrut}^{\mstrut #1}}
\def\downupindices#1#2{{\mstrut}_{\mstrut #1}^{\hp{#1}\mstrut #2}}
\def\updownindices#1#2{{\mstrut}^{\mstrut #1}_{\hp{#1}\mstrut #2}}

\def\index#1{{\scriptstyle #1}}

\def\Parder#1#2{
\mathchoice{\partial{#1} \over\partial{#2}}{\partial{#1}/\partial{#2}}{}{} }
\def\parder#1{\partial/\partial{#1}}

\def\L#1{L\downindex{#1}}
\def\E{{\cal E}}
\def\J#1{J\downindex{#1}}
\def\Q#1{Q\downindex{#1}}
\def\C{C}
\def\B#1{{\tilde B}\downindex{#1}}
\def\T#1{\Theta\downindex{#1}}
\def\w#1{\omega\downindex{#1}}
\def\btT#1{\psi\downindex{#1}}
\def\dyn{{}_{\cal D}}
\def\nondyn{{}_{\cal N}}
\def\bc{{\cal F}}
\def\H#1{H\upindex{\rm #1}}
\def\bcdata#1#2{{\cal F}\mixedindices{#2}{#1}}
\def\othbcdata#1#2{\hat{\cal F}\mixedindices{#2}{#1}}
\def\HE{{\cal E}_H}
\def\Hdens{{\cal H}}

\def\region{{\cal V}}
\def\bcsurface{{\cal B}}

\def\tfvec#1{\xi\upindex{#1}}
\def\tfduvec#1{\xi\downindex{#1}}
\def\perptfvec#1{\zeta\upindex{#1}}
\def\tantfvec#1{\xi\mixedindices{#1}{\parallel}}

\def\der#1{\partial\downindex{#1}}
\def\coder#1{\partial\upindex{#1}}
\def\nder#1#2{\partial\mixedindices{#1}{#2}}
\def\D#1{{\cal D}\downindex{#1}}
\def\coD#1{{\cal D}\upindex{#1}}
\def\covder#1{\nabla\downindex{#1}}
\def\covcoder#1{\nabla\upindex{#1}}
\def\covSder#1{\nabla\mixedindices{S}{#1}}
\def\covcoSder#1{\nabla\upindex{S #1}}
\def\perpD#1{{\cal D}\mixedindices{\perp}{#1}}
\def\perpcoD#1{{\cal D}\upindex{\perp #1}}
\def\perpcovder#1{\nabla\mixedindices{\perp}{#1}}

\def\covgder#1{{{}^g\nabla}\downindex{#1}}
\def\covgcoder#1{{{}^g\nabla}\upindex{#1}}
\def\div#1{\partial\mixedindices{#1}{\Sigma}}
\def\codiv#1{\partial\mixedindices{\Sigma}{#1}}

\def\g#1#2{g\downupindices{#1}{#2}}
\def\metric#1#2{\sigma\downupindices{#1}{#2}}
\def\perpmetric#1#2{\sigma\mixedindices{\perp}{#1}\upindex{#2}}
\def\flat#1#2{\eta\downupindices{#1}{#2}}
\def\vol#1#2{\epsilon\downupindices{#1}{#2}}
\def\duvol#1#2{{*\epsilon}\downupindices{#1}{#2}}
\def\perpvol#1#2{\epsilon\mixedindices{\perp}{#1}\upindex{#2}}
\def\coordvol#1#2{\varepsilon\downupindices{#1}{#2}}
\def\e#1#2{e\downupindices{#1}{#2}}
\def\inve#1#2{e\updownindices{#1}{#2}}
\def\x#1#2{x\mixedindices{#1}{#2}}
\def\id#1#2{\delta\mixedindices{#2}{#1}}

\def\curv#1#2{R\downupindices{#1}{#2}}
\def\scurv{R}
\def\con#1#2{\Gamma\updownindices{#1}{#2}}
\def\tcurv#1#2{{\tilde R}\downupindices{#1}{#2}}
\def\curvS#1#2{{\cal R}\downupindices{#1}{#2}}
\def\scurvS{{\cal R}}
\def\excurv#1#2{\kappa\mixedindices{#1}{#2}}
\def\trexcurv{\kappa}
\def\norScon#1#2{{\cal J}\mixedindices{#1\perp}{#2}}
\def\norScurv#1#2{{\cal R}\mixedindices{\perp}{#1}\upindex{#2}}
\def\perpcon#1#2{J\mixedindices{#1\perp}{#2}}
\def\perpcurv#1#2{R\mixedindices{\perp}{#1}\upindex{#2}}

\def\surfcurv#1#2{{\cal R}\downupindices{#1}{#2}}
\def\surfexcurv#1#2{{\cal K}\downupindices{#1}{#2}}
\def\surfder#1#2{{\cal D}\mixedindices{#2}{#1}}
\def\surfaccel#1{a\downindex{#1}}

\def\EMT#1#2{T\downupindices{#1}{#2}}
\def\G#1#2{G\downupindices{#1}{#2}}

\def\a#1#2{\alpha\mixedindices{#1}{#2}}
\def\ta#1#2{\tilde\alpha\mixedindices{#1}{#2}}
\def\b#1#2{\beta\mixedindices{#1}{#2}}
\def\tb#1#2{\tilde\beta\mixedindices{#1}{#2}}
\def\asub#1#2#3{\alpha_{#1}\mixedindices{#2}{#3}}

\def\s#1#2{s\mixedindices{#1}{#2}}
\def\t#1#2{t\mixedindices{#1}{#2}}
\def\h#1#2{h\downupindices{#1}{#2}}
\def\K#1#2{K\downupindices{#1}{#2}}
\def\trK{\K{}{}}
\def\N#1#2{N\mixedindices{#1}{#2}}
\def\tN#1#2{{\cal N}\mixedindices{#1}{#2}}

\def\n#1#2{n\mixedindices{#1}{#2}}
\def\q#1#2{q\downupindices{#1}{#2}}
\def\p#1#2{p\downupindices{#1}{#2}}
\def\u#1#2{u\mixedindices{#1}{#2}}

\def\svec{\vec s}

\def\news#1#2{s\mixedindices{#1}{#2}}
\def\newt#1#2{t\mixedindices{#1}{#2}}
\def\newh#1#2{h\downupindices{#1}{#2}}
\def\newK#1#2{K\downupindices{#1}{#2}}
\def\newtrK{\K{}{}}

\def\coeff#1#2#3{\Pi_{#1}\mixedindices{#3}{#2}}
\def\othcoeff#1#2#3{\hat\Pi_{#1}\mixedindices{#3}{#2}}

\def\frame#1#2{\theta\mixedindices{#1}{#2}}
\def\outnframe#1#2{\theta^+\mixedindices{#1}{#2}}
\def\innframe#1#2{\theta^-\mixedindices{#1}{#2}}
\def\nframe#1#2{\theta\mixedindices{#1}{#2}}
\def\fixinframe#1#2{\hat\theta{}^+\mixedindices{#1}{#2}}
\def\fixoutframe#1#2{\hat\theta{}^-\mixedindices{#1}{#2}}

\def\frcon#1#2{\Gamma\downupindices{#1}{#2}}
\def\tfrcon#1#2{{\tilde \Gamma}\downupindices{#1}{#2}}
\def\adfr#1#2{\vartheta\mixedindices{#1}{#2}}
\def\adothfr#1#2{\tilde\vartheta\mixedindices{#1}{#2}}

\def\U#1#2{U\updownindices{#1}{#2}}

\def\mcurvH#1#2{H\mixedindices{#1}{#2}}
\def\perpmcurvH#1#2{H\mixedindices{#1}{\perp #2}}
\def\Pperp#1#2#3{(P^{#1}_\perp)\mixedindices{#2}{#3}}
\def\Ppar#1#2#3{(P^{#1}_\parallel)\mixedindices{#2}{#3}}
\def\normH{|\H{}{}|}
\def\normPperp{|P_\perp|}
\def\hatH#1#2{\hat H\mixedindices{#1}{#2}}
\def\hatperpH#1#2{\hat H\mixedindices{#1}{\perp #2}}
\def\hatP#1#2{\hat \zeta\mixedindices{#1}{#2}}
\def\fixPpar#1#2{{\hat P}_\parallel\mixedindices{#1}{#2}}
\def\fixPperp#1#2{{\hat P}_\perp\mixedindices{#1}{#2}}
\def\fixP#1#2{{\hat P}\mixedindices{#1}{#2}}
\def\P#1#2{P\mixedindices{#1}{#2}}
\def\barP#1#2{\bar P\mixedindices{#1}{#2}}

\def\bs#1#2{s'\mixedindices{#1}{#2}}
\def\bt#1#2{t'\mixedindices{#1}{#2}}
\def\ha{\chi}

\def\fixs#1#2{\hat s\mixedindices{#1}{#2}}
\def\fixt#1#2{\hat t\mixedindices{#1}{#2}}

\def\fixu#1#2{\hat u\mixedindices{#1}{#2}}
\def\fixv#1#2{\hat v\mixedindices{#1}{#2}}

\def\v#1#2{v\mixedindices{#1}{#2}}

\def\PrS{{\cal P}_S}
\def\PrperpS{{\cal P}_S^\perp}
\def\PrSop#1#2{(\PrS)\downupindices{#1}{#2}}
\def\PrperpSop#1#2{(\PrperpS)\downupindices{#1}{#2}}
\def\Pr#1{{\cal P}_{#1}}
\def\Prop#1#2#3{({\cal P}_{#1})\downupindices{#2}{#3}}
\def\Prperp#1{{\cal P}^\perp_{#1}}
\def\Prperpop#1#2#3{({\cal P}^\perp_{#1})\downupindices{#2}{#3}}

\def\ncongruenceS{S_{(\lambda_+,\lambda_-)}}

\def\TS{T(S)}
\def\TperpS{T(S)^\perp}
\def\TM{T(M)}

\def\Lie#1{{\cal L}_{#1}}

\def\O{O}
\def\spi{\iota^0}
\def\scri{{{\cal I}^\pm}}


\def\A#1{A\downindex{#1}}
\def\F#1{F\downindex{#1}}
\def\duF#1{{*F}\downindex{#1}}
\def\coF#1{F\upindex{#1}}

\def\Efield#1{E\downindex{#1}}
\def\Bfield#1{B\downindex{#1}}

\def\Evec#1{{\vec E}_{#1}}
\def\Bvec#1{{\vec B}_{#1}}
\def\Avec#1{{\vec A}_{#1}}
\def\dervec#1{\vec\partial\downindex{#1}}


\def\ymA#1#2{A\mixedindices{\mit#1}{#2}}
\def\ymF#1#2{F\mixedindices{\mit#1}{#2}}
\def\ymcoF#1#2{F\upindex{{\mit#1}#2}}
\def\ymduF#1#2{{*F}\mixedindices{\mit#1}{#2}}
\def\C#1#2{C\downupindices{\mit#1}{\mit#2}}
\def\k#1#2{k\downupindices{\mit#1}{\mit#2}}
\def\duC#1#2{C^*\updownindices{\mit#1}{\mit#2}}

\def\ymconn#1#2#3{\Gamma\mixedindices{\mit#2}{#1}\downindex{\mit#3}}
\def\ymcoconn#1#2#3{\Gamma\upindex{#1{\mit#2}}\downindex{\mit#3}}

\def\W#1#2{W\mixedindices{\mit#1}{#2}}
\def\coW#1#2{W\upindex{{\mit#1}#2}}

\def\ymU#1#2{U\updownindices{\mit#1}{\mit#2}}
\def\yminvU#1#2{U\updownindices{-1{\mit#1}}{\mit#2}}


\def\kg#1{\varphi^{\mit#1}}

\def\ie/{i.e.}
\def\eg/{e.g.}
\def\cf/{c.f.}
\def\const{{\rm const}}

\hyphenation{
}

\def\onfr/{orthonormal frame}
\def\bdc/{boundary condition}
\def\bdt/{boundary term}
\def\af/{asymptotically flat}
\def\KG/{Klein-Gordon}
\def\YM/{Yang-Mills}


\title{Covariant Hamiltonian boundary conditions\\
in General Relativity 
for spatially bounded spacetime regions.}

\author{Stephen C. Anco
\thanks{Email address : sanco@brocku.ca}}

\address{Department of Mathematics, Brock University,\\
St Catharines, Ontario L2S 3A1, Canada}

\author{Roh S. Tung
\thanks{Email address : roh@gr.uchicago.edu}}

\address{Enrico Fermi Institute, University of Chicago,\\
Chicago, Illinois 60637, USA  }

\date{\today}

\maketitle

\begin{abstract}
We investigate the covariant Hamiltonian symplectic structure 
of General Relativity for spatially bounded regions of spacetime
with a fixed time-flow vector field. 
For existence of a well-defined Hamiltonian variational principle
taking into account a spatial boundary,  
it is necessary to modify the standard Arnowitt-Deser-Misner Hamiltonian 
by adding a boundary term
whose form depends on the spatial boundary conditions
for the gravitational field. 
The most general mathematically allowed 
boundary conditions and corresponding boundary terms 
are shown to be determined by solving a certain equation 
obtained from the symplectic current 
pulled back to the hypersurface boundary of the spacetime region. 
A main result is that we obtain a covariant derivation of 
Dirichlet, Neumann, and mixed type boundary conditions
on the gravitational field at a fixed boundary hypersurface,
together with the associated Hamiltonian boundary terms. 
As well, we establish uniqueness of these boundary conditions
under certain assumptions motivated by the form of the symplectic current. 
Our analysis uses a Noether charge method 
which extends and unifies several results developed in recent literature
for General Relativity. 
As an illustration of the method,
we apply it to the Maxwell field equations to derive 
allowed boundary conditions and boundary terms 
for existence of a well-defined Hamiltonian variational principle
for an electromagnetic field in a fixed spatially bounded region of 
Minkowski spacetime. 
\end{abstract}



\section{Introduction}

The mathematical structure of General Relativity 
as a Hamiltonian field theory is well-understood
for asymptotically flat spacetimes. 
As first shown by Regge and Teitelboim 
\cite{Regge-Teitelboim},
with asymptotic fall-off conditions on the metric, 
there is a modification of the standard 
Arnowitt-Deser-Misner (ADM) Hamiltonian 
\cite{ADM}
whose field equations obtained from 
the Hamiltonian variational principle
yield a 3+1 split of the Einstein equations. 
The ADM Hamiltonian itself yields 
the 3+1 Einstein equations 
only if compact support variations of the metric
are used in the variational principle. 
For metric variations satisfying asymptotic fall-off conditions,
the ADM Hamiltonian 
does not give a well-defined variational principle 
since its variation produces 
asymptotic boundary terms that do not vanish. 
However, 
the boundary terms can be canceled 
by the addition of a surface integral term at spatial infinity
to the ADM Hamiltonian. 
The resulting Regge-Teitelboim Hamiltonian 
yields a well-defined variational principle 
for the Einstein equations
with asymptotic fall-off conditions on the metric. 
On solutions of the Einstein equations
the Hamiltonian reduces to a surface integral expression 
over spatial infinity,
which turns out to yield the physically important
ADM definition of total energy, momentum, and angular momentum
for asymptotically flat spacetimes. 
Various modern, covariant formulations of 
this Hamiltonian structure are known
\cite{Witten,Wald-Lee,GIMSY,Nester,Kijowski}.

A natural question to investigate 
is whether this Hamiltonian structure 
can be extended to spatially bounded regions of spacetime. 
An important motivation is astrophysical applications 
where asymptotically flat boundary conditions are not appropriate,
e.g. collapse to a black-hole, mergers of binary stars, 
or collision of black-holes. 
Another important application is for numerical solution methods
of the Einstein equations. 
In these situations the spatial boundary 
is not an actual physical boundary in spacetime,
but rather is viewed as a mathematically defined timelike hypersurface 
whose boundary conditions effectively replace 
the dynamics of the gravitational field in the exterior region.

In this paper and a sequel \cite{symplecticvectors}, 
we work out the covariant Hamiltonian structure of General Relativity
for arbitrary spatially compact regions of spacetime
$\Sigma\times\Rnum$ whose spacelike slices possess
a closed two-surface boundary $\partial\Sigma$, 
with a fixed time-flow vector field tangent 
to the timelike hypersurface boundary $\partial\Sigma\times\Rnum$.
Rather than start with given boundary conditions on the metric,
we instead seek to determine both 
the most general surface integral term 
necessary to be added to the ADM Hamiltonian in covariant form 
together with the most general corresponding boundary conditions 
on the metric at $\partial\Sigma$
such that the modified Hamiltonian has well-defined
variational derivatives. 
This would yield the most general mathematically allowed 
variational principle for the Einstein equations 
with spatial boundary conditions on the metric. 
To carry out the analysis
we employ the covariant Hamiltonian formalism 
(referred to as the Noether charge method) 
developed in 
\Ref{Wald-Lee,Wald-Iyer1,Wald-Zoupas}. 

The main results are that 
we find Dirichlet and Neumann type boundary conditions
for the metric at a spatial boundary two-surface
and obtain the associated Hamiltonian surface integrals. 
Under some natural assumptions motivated by 
the symplectic structure arising from the ADM Hamiltonian,
the most general allowed boundary conditions are shown to be
certain types of mixtures of the Dirichlet and Neumann ones. 
We also investigate the geometrical structure of
the Dirichlet and Neumann covariant Hamiltonians. 
These each turn out to involve an underlying 
``energy-momentum'' vector at each point in the tangent space 
of the spacetime at the two-surface. 
In the Dirichlet case, 
this vector depends only on 
the extrinsic geometry of the spatial boundary two-surface. 
Most strikingly, 
when the vector is decomposed into tangential and normal parts
with respect to the two-surface, 
the normal part yields a direction in which 
the two-surface has zero expansion in the spacetime. 

In \secref{maxwell}
we first apply the Noether charge formalism
to investigate, as an illustrative example,
the covariant Hamiltonian structure of the free Maxwell equations
on spatially compact regions of Minkowski spacetime. 
We show that this analysis leads to 
Dirichlet and Neumann type boundary conditions 
on the electromagnetic field, 
corresponding to conductor and insulator type boundaries,
as well as mixed type boundary conditions
which are linear combinations of 
the Dirichlet and Neumann ones. 
We also investigate more general boundary conditions 
which give rise to a well-defined Hamiltonian variational principle
for the Maxwell equations, 
and we obtain a uniqueness result for 
the mixed and pure type Dirichlet and Neumann boundary conditions 
under some assumptions. 
The associated Dirichlet and Neumann Hamiltonians 
are shown to reduce to expressions for the total energy
of the electromagnetic field,
including contributions from surface electric charge and current
due to the boundary conditions. 

In \secref{analysis}
we carry out the corresponding analysis of 
the covariant Hamiltonian structure of General Relativity
for arbitrary spatially compact regions of spacetime
with a closed two-surface boundary,
without matter fields. 
We make some concluding remarks in \secref{conclusion}. 
In an Appendix we develop the Noether charge method 
for general Lagrangian field theories with a time symmetry.
This approach extends and unifies some aspects of 
the covariant Hamiltonian formalism introduced in recent literature
\cite{Wald-Lee,Wald-Iyer1,Wald-Zoupas,Wald-Iyer2}. 
(Throughout we use the notation and conventions of 
\Ref{Wald-book}.)

Inclusion of matter fields and  
analysis of the geometrical properties of 
the resulting Dirichlet and Neumann Hamiltonians 
for General Relativity will be investigated in \Ref{symplecticvectors}. 
It is also left to that paper to discuss
the relation between these Hamiltonians 
and the Regge-Teitelboim Hamiltonian 
in the case when the two-surface boundary is taken 
in a limit to be spatial infinity in an asymptotically flat spacetime.

\section{Electrodynamics}
\label{maxwell}

To illustrate our basic approach 
and the covariant Hamiltonian (Noether charge) formalism,
we consider the free Maxwell field theory 
in 4-dimensional Minkowski spacetime
$(R^4,\flat{ab}{})$.  

We use the standard electromagnetic field Lagrangian, 
where the field variable is the electromagnetic potential 1-form $\A{a}$,
with the field strength 2-form defined as $\F{ab}=\der{[a} \A{b]}$. 
The Lagrangian 4-form for the field $\A{a}$ is given by
\EQ\label{MEL}
\L{abcd}(A)
=\frac{1}{2} \vol{abcd}{} \F{mn} \coF{mn}
= 3\F{[ab} \duF{cd]}
\endEQ
where $\duF{ab}=\vol{ab}{cd}\F{cd}$ 
is the dual field strength 2-form
defined using the volume form $\vol{abcd}{}$. 
A variation of this Lagrangian gives
\EQ\label{MEvarL}
\delta \L{abcd}(A)
= 
\der{[a}\T{bcd]}(A,\delta A)
+6\delta\A{[a}( \der{b}\duF{cd]} )
\endEQ
where 
\EQ\label{MET}
\T{bcd}(A,\delta A) 
=6 \delta\A{[b}\duF{cd]} 
\endEQ
defines the symplectic potential 3-form. 
From \Eqref{MEvarL},
one obtains the field equations 
\EQ\label{AME}
\E\downindex{bcd}(A)= 6 \der{[b}\duF{cd]} =0 , 
\endEQ
or equivalently, after contraction with the volume form, 
\EQ
\coder{a} \F{ab}=\coder{a} \der{[a} \A{b]} =0
\endEQ
which is the source-free Maxwell equations for $\A{a}$. 

Let $\tfvec{a}=(\parder{t})^a$ 
be a timelike isometry of the Minkowski metric,
with unit normalization $\tfvec{a} \tfvec{b} \flat{ab}{}=-1$,
and let $\Sigma$ be a region contained in a spacelike hyperplane $t=0$
orthogonal to $\tfvec{a}$
with the boundary of the region being 
a closed 2-surface $\partial\Sigma$. 
Denote the unit outward spacelike normal to $\partial\Sigma$ in $\Sigma$
by $\s{a}{}$, 
and denote the metric and volume form on $\partial\Sigma$
by $\metric{ab}{}=\flat{ab}{} -\s{}{a}\s{}{b}+\tfduvec{a}\tfduvec{b}$ 
and $\vol{ab}{}=\vol{abcd}{} \s{c}{} \tfvec{d}$. 
Let $\Sigma_t$ and $\partial\Sigma_t$ be the images of 
$\Sigma$ and $\partial\Sigma$ under the one-parameter diffeomorphism 
generated by $\tfvec{a}$ on Minkowski spacetime. 
Denote the metric compatible derivative operator on $\partial\Sigma$ 
by $\D{a}$. 
Let 
\EQ\label{MEliederA}
\Lie{\xi} \A{a} 
= \tfvec{e} \der{e} \A{a} +\A{e}\der{a} \tfvec{e}
= 2\tfvec{e} \der{[e} \A{a]} +\der{a}( \tfvec{e} \A{e} ) ,
\endEQ
which is the Lie derivative of $\A{a}$
with respect to $\tfvec{e}$.

The Noether current 3-form associated to $\tfvec{a}$ is given by
\EQ\label{MEJ}
\J{abc}(\xi,A)
=\T{abc}(A,\Lie{\xi}A)+ 4\tfvec{d}\L{abcd}(A)
=6 \duF{[bc} \Lie{\xi}\A{a]}  
+12 \tfvec{d} \F{[ab}\duF{cd]} , 
\endEQ
which simplifies to
\EQs
\J{abc}(\xi,A)
&&
= 6 \der{[a}( \duF{bc]}\tfvec{d}\A{d} ) 
+2 \tfvec{e} \vol{abcd}{} ( \id{e}{d} \F{mn} \coF{mn} -4 \F{en} \coF{dn} )
-\tfvec{e} \A{e} \E\downindex{abc}(A)
\label{MEsimpleJ}
\endEQs
after use of \Eqrefs{AME}{MEliederA}. 
Hence, one obtains the Noether current on solutions $\A{a}$, 
\EQ
\J{abc}(\xi,A)
=  6 \der{[a}( \duF{bc]}\tfvec{d}\A{d} ) 
+2 \tfvec{e} \vol{abcd}{} ( \id{e}{d} \F{mn} \coF{mn} -4 \F{en} \coF{dn} ) . 
\endEQ
(Note, one easily sees that this 3-form $\J{abc}(\xi,A)$ 
is closed but is not exact,
i.e. there does not exist a Noether current potential 
$\Q{bc}(\xi,A)$ satisfying
$\J{abc}(\xi,A)=3\der{[a}\Q{bc]}(\xi,A)$.)
Correspondingly, the Noether charge on solutions $\A{a}$ is given by 
\EQ
Q_\Sigma(\xi)
= \int_\Sigma \J{abc}(\xi;A)
= \int_\Sigma \vol{abcd}{} \tfvec{e}( 
-8\F{en} \coF{dn} +2 \id{e}{d} \F{mn} \coF{mn} )
+\oint_{\partial\Sigma} 2 \duF{bc}\tfvec{d}\A{d} . 
\endEQ
This expression simplifies in terms of 
the electromagnetic stress-energy tensor defined by 
\EQ
\EMT{e}{d}(F) 
= 2\F{en} \coF{dn} -\frac{1}{2} \id{e}{d} \F{mn} \coF{mn} . 
\endEQ
Thus, 
\EQs
\frac{1}{4} Q_\Sigma(\xi)
&&
= \int_\Sigma \tfvec{e}\tfvec{d} \EMT{de}{}(F) d^3x
+ \oint_{\partial\Sigma} \tfvec{a} \A{a} \tfvec{d} \s{e}{} \F{de} dS
\label{MEQ}
\endEQs
where $d^3x$ and $dS$ denote the coordinate volume elements on 
$\Sigma$ and $\partial\Sigma$
obtained from the volume forms 
$\vol{abcd}{}\tfvec{d}$ and $\vol{bc}{}$, 
respectively.

\subsection{Covariant Hamiltonian formulation}
\label{MEpreliminaries}

The symplectic current, 
defined by the antisymmetrized variation of $\T{bcd}(A,\delta A)$,
is given by the 3-form 
\EQ\label{MEw}
\frac{1}{6} \w{bcd}(\delta_1 A, \delta_2 A)
= \delta_1\A{[b}\delta_2\duF{cd]} 
- \delta_2\A{[b}\delta_1\duF{cd]} . 
\endEQ
Then the presymplectic form on $\Sigma$ is given by 
\EQ\label{MEsymplecticform}
\Omega_\Sigma(\delta_1 A,\delta_2 A) 
= \int_\Sigma \w{bcd}(\delta_1 A,\delta_2 A) . 
\endEQ
A Hamiltonian conjugate to $\xi$ on $\Sigma$ 
is a function $H_\Sigma(\xi;A)=\int_\Sigma \Hdens\downindex{abc}(\xi;A)$
for some locally constructed 3-form $\Hdens\downindex{abc}(\xi;A)$
such that
\EQ\label{MEHdef}
\delta H_\Sigma(\xi;A) \equiv H'_\Sigma(\xi;A,\delta A)
= \Omega_\Sigma(\delta A, \Lie{\xi}A)
\endEQ
for arbitrary variations $\delta\A{a}$ away from solutions $\A{a}$. 

From the expression \eqref{MEJ} for the Noether current,
the presymplectic form yields
\EQs
\Omega_\Sigma(\delta A, \Lie{\xi}A)
&& 
= \int_\Sigma \delta \J{abc}(\xi,A) 
+4 \tfvec{d} \delta\A{[d} \E\downindex{abc]}(A) 
-\oint_{\partial \Sigma} \tfvec{c} \T{abc}(A,\delta A) . 
\label{MEpresymplectic}
\endEQs
Hence, 
for compact support variations $\delta\A{a}$ away from solutions $\A{a}$, 
the Noether current gives a Hamiltonian \eqref{MEHdef}
with $\Hdens\downindex{abc}(\xi;A) =\J{abc}(\xi,A)$, 
up to an inessential boundary term. 
The simplified expression \eqref{MEsimpleJ} for $\J{abc}(\xi,A)$
thereby yields the Hamiltonian 
\EQ\label{MEhamiltonian}
H(\xi;A)
= 4 \int_\Sigma  
\tfvec{e}\tfvec{d}( \EMT{de}{}(F) +\A{e}\coder{c}\F{cd} )d^3x .
\endEQ
On solutions $\A{a}$, this Hamiltonian is equal to 
the total electromagnetic field energy on $\Sigma$, 
$H(\xi;A) = 4 \int_\Sigma  \tfvec{e}\tfvec{d} \EMT{de}{}(F) d^3x$. 

To define a Hamiltonian \eqref{MEHdef}
for variations $\delta\A{a}$ without compact support,
it follows that the term 
$\tfvec{c}  \T{abc}(A, \delta A)$ 
in \Eqref{MEpresymplectic}
needs to be a total variation at the boundary $\partial \Sigma$,
i.e. there must exist a locally constructed 3-form 
$\B{abc}(A)$ such that one has
\EQ\label{MEbceq}
\tfvec{c} \T{abc}(A, \delta A) |_{\partial\Sigma}
=( \tfvec{c} \delta\B{abc}(A) 
+\der{[a}\a{}{b]}(\xi;A, \delta A) )|_{\partial\Sigma}
\endEQ
where $\a{}{b}(\xi;A, \delta A)$ is a locally constructed 1-form.
This equation holds if and only if, 
by taking an antisymmetrized variation \cite{Wald-Zoupas},
one has
\EQ\label{MEbcskeweq}
\vol{}{bc}\tfvec{a} 
\w{abc}(\delta_1 A,\delta_2 A)|_{\partial\Sigma}
=\D{c}\tb{c}{}(\xi;\delta_1 A,\delta_2 A)|_{\partial\Sigma}
\endEQ
where
\EQ\label{MEbeq}
\tb{c}{}(\xi;\delta_1 A,\delta_2 A)
=\vol{}{cb} \delta_1\a{}{b}(\xi;A,\delta_2 A)
- \vol{}{cb} \delta_2\a{}{b}(\xi;A,\delta_1 A)
\endEQ 
is a locally constructed vector, in $T(\partial\Sigma)$,
which is skew bilinear in $\delta_1 A,\delta_2 A$. 
The term involving the symplectic current is given by 
\EQs
\vol{}{bc}\tfvec{a} \w{abc}(\delta_1 A,\delta_2 A)
&&
= 8\s{c}{} \h{}{de}
( \delta_1\A{e} \delta_2\F{cd} - \delta_2\A{e} \delta_1\F{cd} )
\endEQs
with 
$\h{ab}{} =\flat{ab}{} - \s{}{a}\s{}{b} 
= \metric{ab}{} -\tfduvec{a}\tfduvec{b}$. 
Given a solution of equation \eqref{MEbcskeweq}, 
one can then determine $\B{abc}(A)$ from equation \eqref{MEbceq} by
\EQs
\vol{}{ab} \tfvec{c} \B{abc}(A) &&
= \vol{}{ab}( \tfvec{c}\T{abc}(A,\delta A) -\der{a}\a{}{b}(\xi;A,\delta A) )
\nonumber\\&&
= 8 \s{c}{} \h{}{de} \F{cd}\delta\A{e} -\D{a}\ta{a}{}(\xi;A,\delta A)
\label{MEBeq}
\endEQs
where
\EQ\label{MEtaeq}
\ta{a}{}(\xi;A,\delta A) = \vol{}{ab}\a{}{b}(\xi;A,\delta A) .
\endEQ
This leads to the following main result. 

\Proclaim{ Proposition 2.1. }{
A Hamiltonian conjugate to $\tfvec{a}$ on $\Sigma$
exists for variations $\delta\A{a}$ with support on $\partial\Sigma$
if and only if 
\EQ\label{MEbcdeteq}
8 \h{}{bc}\s{a}{}( 
\delta_1\A{b} \der{[a}\delta_2\A{c]} 
- \delta_2\A{b} \der{[a}\delta_1\A{c]} )|_{\partial\Sigma}
=\D{c}\tb{c}{}(\xi;\delta_1 A,\delta_2 A)
\endEQ
for some locally constructed vector $\tb{}{a}(\xi,\delta_1 A,\delta_2 A)$,
in $T(\partial\Sigma)$, which is skew bilinear in $\delta_1 A,\delta_2 A$. 
The solutions of equation \eqref{MEbcdeteq} of the form
$\delta\bcdata{a}{}(A)|_{\partial\Sigma}=0$
give the allowed boundary conditions 
$\bcdata{a}{}(A)|_{\partial\Sigma}$
for a Hamiltonian formulation of the Maxwell equations 
in the local spacetime region $\Sigma_t$, $t\geq 0$. 
For each boundary condition, 
there is a corresponding Hamiltonian 
given by the Noether charge plus a boundary term
\EQ\label{MEH}
H_\Sigma(\xi;A)
= \int_\Sigma \J{abc}(\xi,A) -\oint_{\partial\Sigma} \tfvec{a}\B{abc}(A)
\equiv H(\xi;A) + H_B(\xi;A) , 
\endEQ
with 
\EQs
&&
H(\xi;A)
= 4 \int_\Sigma \tfvec{e}\tfvec{d} 
( \EMT{de}{}(F) +\A{e}\coder{c} \F{cd} )d^3x , 
\\&&
H_B(\xi;A)
= \oint_{\partial\Sigma}  \tfvec{a}
( 4 \A{a} \tfvec{d} \s{e}{} \F{de} - \frac{1}{2}\B{a}(A) ) dS , 
\endEQs
where $\B{a}(A)= \vol{}{bc} \B{abc}(A)$ 
is determined from 
\EQ\label{MEBdeteq}
( \tfvec{a} \B{a}(A) -8 \s{c}{} \h{}{de} \F{cd}\delta\A{e} )|_{\partial\Sigma} 
= \D{a}\ta{a}{}(\xi;A,\delta A) |_{\partial\Sigma}
\endEQ 
with $\ta{a}{}(\xi;A,\delta A)$ given by \Eqrefs{MEbeq}{MEtaeq}. 
Note, $\B{a}(A)$ is unique up to addition of 
an arbitrary covector function of the fixed boundary data $\bcdata{a}{}(A)$. 
}

The results in Proposition~2.1 
take a more familiar form when expressed in terms of 
the electric and magnetic fields on $\Sigma$ defined by 
$\Efield{a}=2\tfvec{b} \F{ab}$, 
$\Bfield{a}=\tfvec{b} \duF{ab}$,
which are vectors in $T(\Sigma)$
(\ie/ $\tfvec{a} \Efield{a}= \tfvec{a} \Bfield{a} =0$). 
A convenient notation now is to write 
vectors in $T(\Sigma)$ using an over script $\rightarrow$,
and for tensors on $M$, 
to denote tangential and normal components with respect to $\Sigma$
by subscripts $\parallel$ and $\perp$,
and denote components orthogonal to $\Sigma_t$ 
by a subscript $0$. 
Then we have 
\EQ
\Evec{} = \dervec{}\A{0} -\der{0}\Avec{} ,\quad 
\Bvec{} = \dervec{} \times \Avec{} . 
\endEQ
In this notation, 
the presymplectic form \eqref{MEsymplecticform}
and the Hamiltonian \eqref{MEH}
reduce to the expressions
\EQs
\frac{1}{4} \Omega_\Sigma(\delta_1 A, \delta_2 A)
&&
= \int_\Sigma 
( \delta_2\Avec{} \cdot\delta_1\Evec{} 
-  \delta_1\Avec{} \cdot\delta_2\Evec{} ) d^3x
\nonumber\\&&
=  \int_\Sigma 
( \delta_1\Avec{} \cdot
( \der{0}\delta_2\Avec{} -\dervec{}\delta_2\A{0} )
-\delta_2\Avec{} \cdot
( \der{0}\delta_1\Avec{} -\dervec{}\delta_1\A{0} )  ) d^3x
\label{MEsymplecticsplit}
\endEQs
and
\EQ
\frac{1}{2} H_\Sigma(\xi;A)
=  \int_\Sigma \frac{1}{2}( \Evec{}{}^2 + \Bvec{}{}^2 )
+\A{0}\dervec{}\cdot \Evec{\perp}\ d^3x
- \oint_{\partial\Sigma} \A{0}\Efield{\perp} + \frac{1}{4}\B{0}(A) dS . 
\endEQ

Note that the Hamiltonian field equations obtained from $H_\Sigma(\xi;A)$
are given by the variational principle
\EQ\label{MEHeqs}
H'_\Sigma(\xi;A,\delta A) 
= \Omega_\Sigma(\delta A, \Lie{\xi}A)
\endEQ
for arbitrary variations $\delta\A{a}|_\Sigma$. 
These field equations split into 
dynamical equations and constraint equations, 
corresponding to a decomposition of $\A{a}$ into 
dynamical and non-dynamical components, respectively $\Avec{}$ and $\A{0}$, 
determined by \cite{Wald-Lee} the degeneracy of 
the presymplectic form \eqref{MEsymplecticsplit}. 
In particular, this yields the Gauss-law constraint equation
\EQ
\dervec{} \cdot \Evec{} 
= \Delta \A{0} -\der{0}  \dervec{} \cdot \Avec{}
=0 
\endEQ
obtained from $H'_\Sigma(\xi;A,\delta \A{0})=0$ 
through variation of $\A{0}$, 
and the dynamical Maxwell evolution equation 
\EQ
\der{0} \Evec{} - \dervec{} \times \Bvec{} 
= (-\der{0}{}^2 +\Delta) \Avec{} 
+\dervec{}( \der{0}\A{0} -\dervec{} \cdot \Avec{} )
=0 
\endEQ
obtained from $H'_\Sigma(\xi;A,\delta \Avec{}) 
= \Omega_\Sigma(\delta \Avec{}, \Lie{\xi}A)$
through variation of $\Avec{}$,
where $\Delta=\dervec{}\cdot\dervec{}$ is the Laplacian on $\Sigma$. 
Thus, the Noether charge (covariant Hamiltonian) formalism 
here is equivalent to the standard canonical formulation \cite{Wald-book}
of the Maxwell equations.

\subsection{Dirichlet and Neumann Boundary Conditions}

Two immediate solutions of 
the determining equation \eqref{MEbcdeteq} with $\tb{a}{}=0$
are boundary conditions 
associated with fixing components of $\A{a}$ or $\F{bc}=\der{[b}\A{c]}$ 
at $\partial\Sigma_t$ for $t\geq 0$. 

Consider 
\EQ\label{MEDbc}
{\rm (D)} \qquad
\metric{b}{a}\delta\A{a}|_{\partial\Sigma_t}=0 ,\quad
\tfvec{a} \delta\A{a}|_{\partial\Sigma_t}=0 ,\quad
t\geq 0, 
\endEQ
or equivalently 
$\delta \Avec{\parallel}=\delta \A{0}=0$, for $t\geq 0$, 
called Dirichlet boundary conditions, 
\ie/ 
\EQ
\bcdata{a}{\rm D}(A)=\h{a}{b}\A{b} ; 
\endEQ
\EQ\label{MENbc}
{\rm (N)} \qquad
\metric{b}{a} \s{c}{}\delta\F{ac}|_{\partial\Sigma_t}=0 ,\quad
\s{a}{} \tfvec{c}\delta\F{ac}|_{\partial\Sigma_t}=0 ,\quad
t\geq 0,
\endEQ
or equivalently 
$\dervec{\perp}\times\delta\Avec{\parallel} 
- \dervec{\parallel}\times\delta\Avec{\perp}
= \dervec{\perp}\delta\A{0} - \der{0}\delta\Avec{\perp}
=0$, for $t\geq 0$, 
called Neumann boundary conditions, 
\ie/
\EQ
\bcdata{a}{\rm N}(A)=\h{a}{b}\s{c}{}\F{bc} . 
\endEQ

\Proclaim{ Theorem 2.2. }{
For the boundary conditions {\rm (D)} or {\rm (N)},
a Hamiltonian conjugate to $\tfvec{a}$ on $\Sigma$,
evaluated on solutions $\A{a}$, is given by 
\EQs
\frac{1}{2} \H{D}(\xi;A)
&&
= 2\int_\Sigma \vol{dabc}{} (
2\tfvec{e} \F{en}\coF{dn} -\frac{1}{2} \tfvec{d} \F{mn}\coF{mn} ) 
+ 2 \oint_{\partial\Sigma} \vol{bc}{} 
\tfvec{a}\A{a} \tfvec{d} \s{e}{} \F{de} 
\\&&
= \frac{1}{2} \int_\Sigma \Evec{}{}^2 + \Bvec{}{}^2 \ d^3x
- \oint_{\partial\Sigma} \A{0}\Efield{\perp} \ dS , 
\label{MEDhamiltonian}\\
\frac{1}{2} \H{N}(\xi;A)
&& 
= 2 \int_\Sigma \vol{dabc}{} (
2\tfvec{e} \F{en}\coF{dn} -\frac{1}{2} \tfvec{d} \F{mn}\coF{mn} ) 
+ 2 \oint_{\partial\Sigma} \vol{bc}{} 
\metric{}{bd} \A{b} \s{e}{} \F{de}
\\&& 
= \frac{1}{2} \int_\Sigma \Evec{}{}^2 + \Bvec{}{}^2\ d^3x
- \oint_{\partial\Sigma} ( \Avec{} \times \Bvec{} )_\perp\ dS . 
\label{MENhamiltonian}
\endEQs }

\proclaim{ Proof: }
For (D), one has
\EQs
\vol{}{bc}\tfvec{a}\T{abc}(A,\Lie{\xi}A)
&& 
= 6 \vol{}{bc}\tfvec{a} \delta\A{[a} \duF{bc]} 
\nonumber\\&&
= 8 ( \tfvec{a}\delta\A{a} \tfvec{d}\s{e}{}\F{de} 
-\metric{}{ad} \delta\A{a} \s{e}{}\F{de} )
= 0
\endEQs
and hence $\B{abc}(A)=0$,
so thus $\tfvec{a}\B{a}(A)=0$. 
For (N), one has
\EQs
\vol{}{bc}\tfvec{a}\T{abc}(A,\Lie{\xi}A)
&&
=  6 \vol{}{bc}\tfvec{a} ( \delta\A{[a} \duF{bc]} ) 
\nonumber\\&& 
=  6 \vol{}{bc}\tfvec{a} ( \delta(\A{[a} \duF{bc]}) - \A{[a} \delta\duF{bc]} )
\nonumber\\&& 
= \delta ( 6\vol{}{bc}\tfvec{a} \A{[a} \duF{bc]} ) 
+8 ( - \tfvec{a} \A{a} \tfvec{d}\s{e}{}\delta\F{de} 
+ \A{a} \metric{}{ad}\s{e}{} \delta \F{de} )
\\&& 
= \delta( \vol{}{bc}\tfvec{a} \B{abc} )
\nonumber
\endEQs
with $\B{abc}(A)=6 \A{[a} \duF{bc]}$. 
Thus,
\EQ
\tfvec{a}\B{a}(A) = 6\vol{}{bc}\tfvec{a} \A{[a} \duF{bc]} 
= 8 \s{}{d} \A{e} \coF{de} . 
\endEQ
\endproof

Note that for both \bdc/s (D) and (N), 
the surface integral terms in the Hamiltonian 
take the form 
\EQ\label{MEP}
H_B(\xi;A) = 4 \oint_{\partial\Sigma} \tfvec{a} \P{}{a}(A) dS
\endEQ
where
\EQs
\P{\rm D}{a}(A)
&& 
= \A{a} \tfvec{b} \s{c}{}\F{bc} ,
\\
\P{\rm N}{a}(A)
&& 
= -\tfduvec{a} \s{d}{} \metric{}{bc} \A{b} \F{cd} . 
\endEQs

There is simple physical interpretation of 
the (D) and (N) boundary conditions:
(D) involves fixing $\A{0}$ and $\Avec{\parallel}$ 
at $\partial\Sigma_t$ for $t\geq 0$,
which means 
\EQ\label{MEconductorbc}
\Evec{\parallel}=\dervec{\parallel}\A{0} -\der{0}\Avec{\parallel} ,\quad 
\Bvec{\perp} = \dervec{\parallel}\times \Avec{\parallel} 
\endEQ
are specified data at the boundary surface 
(analogous to a conductor) as a function of time. 
Note, consequently, 
$\Evec{\perp}$ and $\Bvec{\parallel}$ 
are left free by the boundary condition (D)
and therefore are induced data for solutions $\A{0},\Avec{}$
of the Hamiltonian field equations;
(N) reverses the role of the induced and fixed data
at the boundary surface,
so now $\Evec{\perp}$ and $\Bvec{\parallel}$ 
are specified as a function of time 
(analogous to an insulator),
while the induced data for solutions $\A{0},\Avec{}$
of the Hamiltonian field equations 
are $\Evec{\parallel}$ and $\Bvec{\perp}$. 
Note the fixed data here 
are gauge-equivalent to specifying the normal derivative of 
$\A{0}$ and $\Avec{\parallel}$ 
at $\partial\Sigma_t$ in $\Sigma_t$ for $t\geq 0$, 
\EQ
\Evec{\perp} 
= \dervec{\perp}( \A{0} - \der{0}\chi ) ,\quad 
\Bvec{\parallel} 
= \dervec{\perp}\times ( \Avec{\parallel} -\dervec{\parallel}\chi ), 
\endEQ
where $\chi$ is given by $\dervec{\perp}\chi=\Avec{\perp}$. 

Moreover, 
the Hamiltonians \eqrefs{MEDhamiltonian}{MENhamiltonian}
on solutions $\A{0},\Avec{}$ have the interpretation as 
expressions for the total energy of the electromagnetic fields,
with the surface integral parts representing 
the energy contribution \cite{Jackson} from 
an effective (fictitious) surface charge density in the case (D),
and effective (fictitious) surface current density in the case (N),
associated to the specified data at the boundary surface. 
In particular, 
effective surface charges and currents arise, respectively, 
when $\Evec{\perp}$ or $\Avec{\parallel}$ are left free \cite{Jackson}
on the boundary surface. 

We remark that similar boundary terms arise in the asymptotic case 
when the boundary surface is taken to be a 2-sphere at spatial infinity
on $\Sigma$ (see \Ref{Wald2}).

\subsection{ Determination of allowed boundary conditions }
\label{maxwellbc}

The symplectic current component 
$\vol{}{bc}\tfvec{a} \w{abc}(\delta_1 A,\delta_2 A)$
involves only the field variations
$\h{a}{b}\delta\A{b}$, $\h{a}{b} \s{c}{} \der{[b}\delta\A{c]}$. 
We refer to the components 
$\metric{b}{c}\A{c}, \tfvec{c}\A{c},
\metric{b}{c}\s{a}{} \F{ac}, \tfvec{c}\s{a}{} \F{ac}$, 
or equivalently 
$\Avec{\parallel},\A{0},\Bvec{\parallel},\Evec{\perp}$, 
as {\it symplectic boundary data} at $\partial\Sigma$.
Hence, in solving the determining equation \eqref{MEbcdeteq}
for the allowed boundary conditions on $\A{a}$, 
it is then natural to restrict attention to 
boundary conditions involving only this data. 
(Some remarks on more general boundary conditions are made 
at the end of this section.)
To proceed, 
we suppose that the possible boundary conditions
are linear, homogeneous functions of the symplectic boundary data,
with coefficients locally constructed
out of the geometrical quantities 
$\tfvec{a},\s{a}{},\metric{bc}{},\vol{bc}{}$
at the boundary surface. 
We call this type of boundary condition 
a {\it symplectic} boundary condition. 

\Proclaim{ Theorem 2.3. }{
The most general allowed 
symplectic boundary conditions 
\EQ\label{MEintrinsicbc}
\bcdata{b}{}(\h{c}{d}\A{d},\h{c}{d}\s{e}{}\F{de};
\tfvec{c},\s{c}{},\metric{de}{},\vol{de}{})
\endEQ
for existence of a Hamiltonian conjugate to $\tfvec{a}$ on $\Sigma$
are given by 
\EQ\label{MEbc}
\bcdata{b}{} = 
b_0 \metric{b}{c} \A{c} +a_0 \metric{b}{c} \s{a}{} \F{ac} 
+ b_1 \tfduvec{b}\tfvec{c} \A{c} +a_1 \tfduvec{b}\tfvec{c} \s{a}{}\F{ca} , 
\endEQ
or equivalently, 
\EQs
&& \metric{b}{c} ( 
b_0 \delta\A{c} +a_0 \s{a}{} \delta\F{ac} 
)|_{\partial\Sigma_t} =0, \quad t\geq 0,
\label{mixedbcparallel}\\
&& \tfvec{c}( 
b_1 \delta\A{c} +a_1 \s{a}{}\delta\F{ac} 
)|_{\partial\Sigma_t} =0, \quad t\geq 0
\label{mixedbcperp}
\endEQs
for any constants 
$a_0,b_0$ (not both zero), 
$a_1,b_1$ (not both zero). }

\proclaim{ Proof: }
First, we show that $\tb{a}{}=0$ 
without loss of generality in the determining equation \eqref{MEbcdeteq}
for boundary conditions of the form \eqref{MEintrinsicbc}. 
Note the left side of \Eqref{MEbcdeteq} is algebraic in 
$\delta\A{a},\delta\F{ab}=\der{[a}\delta\A{b]}$. 
Since the right side necessarily involves 
at least one derivative on $\delta\A{a}$, we must have
\EQ\label{tb}
\tb{a}{} =\delta_1\A{b}\delta_2\A{c} \b{abc}{}
\endEQ
for some tensor 
\EQ\label{bhat}
\b{abc}{}=\metric{e}{a}\b{e[bc]}{} . 
\endEQ
We substitute expression \eqref{tb} into \Eqref{MEbcdeteq} 
and collect all terms that do not involve just the symplectic boundary data,
namely 
$\s{c}{}\A{c}$, $\tfvec{c} \metric{d}{a} \F{ac}$, 
$\metric{b}{c}\metric{d}{a} \F{ac}$,
and $\der{(a}\A{b)}$. 
The coefficients of these terms yield algebraic equations
\EQ\label{bhateq1}
\s{}{c} \b{(ab)c}{} =0 ,\quad
\s{}{c} \b{[ab]c}{} =0 ,\quad
\s{[d}{} \b{ab]c}{}=0 . 
\endEQ
Then, since $\b{abc}{}$ has the form \eqref{bhat}, 
we find that the solution of the equations in \eqref{bhateq1} is 
\EQ
\b{abc}{}= 0 
\endEQ
and so $\tb{a}{}=0$. 

Hence, the determining equation \eqref{MEbcdeteq} 
reduces to 
\EQ\label{MEsimpledeteq}
\h{}{bc}\s{a}{}( \delta_1\A{b} \der{[a} \delta_2\A{c]} 
- \delta_1\A{b} \der{[a} \delta_1\A{c]} )|_{\partial\Sigma}
=0 , 
\endEQ
which we are now free to solve 
as a purely algebraic equation in terms of the variables
$\delta\A{b}$ and $\delta\F{ac}=\der{[a}\delta\A{c]}$, 
\ie/
\EQ\label{MEalgebraicdeteq}
\h{}{bc}\s{a}{}( \delta_1\A{b} \delta_2\F{ac} 
- \delta_1\A{b} \delta_2\F{ac} ) 
=0 . 
\endEQ
It is straightforward to show from the form of \Eqref{MEalgebraicdeteq}
that the only solution which is linear, homogeneous 
in the previous variables 
is given by 
\EQ\label{MEbcsol}
\coeff{1}{b}{\ c}\delta\A{c} = \coeff{2}{b}{\ c}\s{a}{}\delta\F{ac}
\endEQ
where $\coeff{1}{}{bc}$ and $\coeff{2}{}{bc}$
are some symmetric tensors orthogonal to $\s{a}{}$, 
such that \Eqref{MEbcsol} can be solved for
either 
$\tfvec{b}\delta\A{b}$ or $\tfvec{b}\s{a}{}\delta\F{ab}$,
and either 
$\metric{c}{b}\delta\A{b}$ or $\metric{c}{b}\s{a}{}\delta\F{ab}$.
Since we require the coefficients in the boundary conditions
under consideration 
to be locally constructed out of 
$\tfvec{a},\s{a}{},\metric{bc}{},\vol{bc}{}$,
we see that
\EQ\label{btens}
\coeff{1}{}{bc}
=b_0\metric{}{bc}+b_1\tfvec{b}\tfvec{c}, \quad
\coeff{2}{}{bc} 
=a_0\metric{}{bc}+a_1\tfvec{b}\tfvec{c}, 
\endEQ
for some constants $a_0,a_1,b_0,b_1$
with $a_0\neq 0$ or $b_0\neq 0$, 
and $a_1\neq 0$ or $b_1\neq 0$. 
This yields the general solution 
\eqrefs{mixedbcparallel}{mixedbcperp}
given in the Theorem. 
\endproof

The boundary conditions given by Theorem~2.3 comprise
the following separate types:
(i) for $a_0=a_1=0$ or $b_0=b_1=0$, 
one obtains, respectively, 
Dirichlet \eqref{MEDbc} and Neumann \eqref{MENbc} boundary conditions;
(ii) for $b_0=0 (a_0\neq 0), b_1\neq 0$, 
the boundary conditions yield a one-parameter $a_1/b_1\equiv c_1$ 
family of the form
\EQ\label{MEothbc}
\metric{b}{c} \s{a}{} \delta \F{ac}{} |_{\partial\Sigma_t} =0 ,\quad
\tfvec{c} \delta\A{c} |_{\partial\Sigma_t} 
= -c_1 \s{a}{} \tfvec{c}\delta \F{ac}{} |_{\partial\Sigma_t} ,\quad
t\ge 0,
\endEQ
or equivalently, 
$\delta\Bvec{\parallel} 
= \svec \delta\A{0} + c_1\frac{1}{2} \delta\Evec{\perp}
=0$, for $t\geq 0$;
(iii) similarly, for $b_1=0 (a_1\neq 0), b_0\neq 0$,
the boundary conditions yield another one-parameter $a_0/b_0\equiv c_0$ 
family of the form
\EQ\label{MEothbc'}
\metric{b}{c} \delta\A{c} |_{\partial\Sigma_t} 
= -c_0 \s{a}{} \delta\F{ab}{} |_{\partial\Sigma_t} ,\quad
\tfvec{c} \s{a}{} \delta\F{ac}{} |_{\partial\Sigma_t}=0 ,\quad
t\ge 0,
\endEQ
or equivalently, 
$\svec\times \delta\Avec{\parallel} +c_0\frac{1}{2} \delta\Bvec{\parallel}
= \delta\Evec{\perp}
=0$, for $t\geq 0$;
(iv) finally, for $b_0\neq 0, b_1\neq 0 (a_0\neq 0, a_1\neq 0)$, 
we obtain a two-parameter $a_1/b_1\equiv c_1$ and $a_0/b_0\equiv c_0$  
family of the boundary conditions 
\EQ\label{MEothbc''}
\metric{b}{c} \delta\A{c} |_{\partial\Sigma_t} 
= -c_0 \s{a}{} \delta\F{ab}{} |_{\partial\Sigma_t} ,\quad
\tfvec{c} \delta\A{c} |_{\partial\Sigma_t} 
= -c_1 \s{a}{} \tfvec{c}\delta \F{ac}{} |_{\partial\Sigma_t} ,\quad
t\ge 0,
\endEQ
or equivalently, 
$\svec\times \delta\Avec{\parallel} +c_0\frac{1}{2} \delta\Bvec{\parallel}
= \svec \delta\A{0} + c_1\frac{1}{2} \delta\Evec{\perp}
=0$, for $t\geq 0$. 

The fixed data for these boundary conditions \eqsref{MEothbc}{MEothbc''}
corresponds to specifying, respectively, the field components
\EQs
&& 
\svec \A{0} + c_1\frac{1}{2} \Evec{\perp} ,\quad
\Bvec{\parallel} ,\quad
c_1\neq 0, 
\\&& 
\Evec{\perp} ,\quad
\svec\times \Avec{\parallel} + c_0\frac{1}{2} \Bvec{\parallel} ,\quad
c_0\neq 0, 
\\&&
\svec \A{0} + c_1\frac{1}{2} \Evec{\perp} ,\quad
\svec\times \Avec{\parallel} + c_0\frac{1}{2} \Bvec{\parallel} ,\quad
c_0\neq 0, c_1\neq 0,
\endEQs
at the boundary surface $\Sigma_t$ for $t\geq 0$.
From Proposition~2.1, we readily obtain the Hamiltonian boundary terms
corresponding to these boundary conditions
(by a proof similar to that for Theorem~2.2). 

\Proclaim{ Theorem 2.4. }{
For boundary conditions \eqsref{MEothbc}{MEothbc''},
there is a respective Hamiltonian \eqref{MEH} 
conjugate to $\tfvec{a}$ on $\Sigma$, 
with boundary terms given by 
\EQs
H_B(\xi;A) && 
= 4\oint_{\partial\Sigma} \vol{bc}{}
( \tfvec{a}\A{a} \tfvec{d} \s{e}{} \F{de} 
+ \metric{}{ad} \A{a} \s{e}{} \F{de} 
-c_1 ( \tfvec{d} \s{e}{} \F{de} )^2 )
\nonumber\\&&
= - \oint_{\partial\Sigma} 
2( \Avec{} \times \Bvec{} )_\perp +c_1 \Evec{\perp}{}^2 \ dS , 
\\
H_B(\xi;A) && 
= 4 \oint_{\partial\Sigma} \vol{bc}{}
c_0 \metric{}{de} \s{a}{} \F{ad} \s{m}{} \F{me} 
\nonumber\\&&
= \oint_{\partial\Sigma} 
c_0 \Bvec{\parallel}{}^2 \ dS , 
\\
H_B(\xi;A) && 
= 4 \oint_{\partial\Sigma} \vol{bc}{}
( \tfvec{a}\A{a} \tfvec{d} \s{e}{} \F{de} 
- c_1 ( \tfvec{d} \s{e}{} \F{de} )^2 
+ c_0 \metric{}{de} \s{m}{} \F{md} \s{n}{} \F{ne} )
\nonumber\\&&
= -\oint_{\partial\Sigma} 
 2\A{0}\Evec{\perp} + c_1 \Efield{\perp}{}^2 -c_0 \Bvec{\parallel}{}^2 \ dS .
\endEQs }

Interestingly, among the allowed boundary conditions given by Theorem~2.3,
we observe that the Hamiltonian boundary terms vanish identically
in one (and only one) case, 
when $c_0=0$ in boundary condition \eqref{MEothbc'}, \ie/ 
\EQ\label{MEspecialbc}
\metric{b}{c} \delta\A{c} |_{\partial\Sigma_t} =0 ,\quad
\tfvec{c} \s{a}{} \delta\F{ac} |_{\partial\Sigma_t} =0 ,\quad 
t\geq 0, 
\endEQ
or equivalently, 
$\delta\Avec{\parallel} =\delta\Evec{\perp}=0$, 
for $t\geq 0$.
The resulting Hamiltonian \eqref{MEH} reduces, on solutions $\A{a}$, 
simply to the expression for the total energy of the electromagnetic fields, 
$H(\xi;A) = 4 \int_\Sigma  \tfvec{e}\tfvec{d} \EMT{de}{}(F) d^3x
= \int_\Sigma \Evec{}{}^2 + \Bvec{}{}^2 \ d^3x$. 
The fixed data corresponding to the boundary condition \eqref{MEspecialbc}
is $\Avec{\parallel}$ and $\Evec{\perp}$, 
which means that the normal components of the electric and magnetic fields
at $\partial\Sigma$ are specified for $t\geq 0$, 
\EQ
\Bvec{\perp} =\dervec{\parallel} \times \Avec{\parallel} ,\quad
\Evec{\perp} = \dervec{\perp}\A{0} -\der{0}\Avec{\perp} . 
\endEQ
Since $\Avec{\parallel}$ and $\Evec{\perp}$ are fixed,
there are no effective charges and currents 
associated to the boundary surface.
Thus, the total electromagnetic energy involves 
no surface integral contributions in this case.

\subsection{Remarks}

We conclude with some short remarks 
on uniqueness of the boundary conditions obtained in Theorem~2.3.

Note that 
the symplectic boundary conditions 
\eqrefs{mixedbcparallel}{mixedbcperp}
are linear combinations of 
the tangential and normal parts of 
the Dirichlet and Neumann boundary conditions,
referred to as mixed boundary conditions. 
In physical terms,
they correspond to specifying 
$b_0 \metric{b}{c} \A{c} +a_0 \s{a}{}\metric{b}{c}\F{ac}$
and 
$b_1 \tfvec{c} \A{c} +a_1 \tfvec{c} \s{a}{}\F{ca}$
as boundary data at $\partial\Sigma_t$ for $t\geq 0$. 
Theorem~2.3 gives a uniqueness result for 
these mixed boundary conditions
under the natural assumption \eqref{MEintrinsicbc} 
about the general type of boundary condition considered
on the fields at the boundary surface. 
If this assumption is loosened,
then there exist additional boundary conditions
allowed by the determining equation \eqref{MEbcdeteq}.

In particular, 
one can trade off some of 
the mixed boundary conditions on the symplectic boundary data
for boundary conditions involving the symmetrized derivatives
of $\A{a}$ at $\partial\Sigma$. 
For example, 
an allowed boundary condition satisfying \Eqref{MEbcdeteq} is 
given by 
$\bc(A)=(
\tfvec{a}\A{a}, 
\metric{}{ab}\der{a}\A{b}, 
\metric{}{ab}\s{c}{}\der{(b}\A{c)}
)$, 
or equivalently
\EQ
\tfvec{a}\delta\A{a}|_{\partial\Sigma_t}=0 ,\quad
\metric{}{ab}\der{a}\delta\A{b}|_{\partial\Sigma_t}=0 ,\quad
\metric{}{ab}\s{c}{}\der{(b}\delta\A{c)}|_{\partial\Sigma_t}=0 ,\quad
t\geq 0,
\endEQ
with $\a{}{a}=8\vol{a}{b}\A{b}\s{c}{}\delta\A{c}$. 
From \Eqref{MEBdeteq} one obtains
\EQ
\frac{1}{8} \tfvec{a} \B{a} 
= \s{e}{} \metric{}{cd}( \A{c} \der{(d}\A{e)} -\A{e} \der{c}\A{d} )
-\frac{1}{2} \A{d}\A{e} \coD{d} \s{e}{}
\endEQ
where $\D{d}\s{}{e}= \metric{d}{m}\metric{e}{n}\der{m}\s{}{n}$
is the extrinsic curvature of $\partial\Sigma$ in $\Sigma$. 
Hence the corresponding boundary term in the Hamiltonian 
is given by \Eqref{MEP} with 
\EQ
\P{}{a}(A) =
\A{a} \s{d}{} \tfvec{e} \F{de}
- \s{e}{} \metric{}{cd}( \A{c} \der{(d}\A{e)} -\A{e} \der{c}\A{d} )
+\frac{1}{2} \A{d}\A{e} \coD{d} \s{e}{} . 
\endEQ

\section{Analysis of General Relativity}
\label{analysis}

We now apply the Noether charge analysis to General Relativity,
specifically to the vacuum Einstein equations for the gravitational field
in a spatially bounded spacetime region
with a fixed time-flow vector field. 
It is straightforward to also include matter fields, 
as we discuss in \Ref{symplecticvectors}.

For General Relativity without matter sources,
the starting point is the standard Lagrangian formulation of
the Einstein equations with the spacetime metric 
as the field variable. 
It turns out, however, that the analysis is considerably simplified
by introduction of a tetrad (orthonormal frame). 
Moreover, 
taking into account local rotations and boosts of the tetrad, 
the boundary conditions and resulting Hamiltonians 
that arise in the tetrad formulation are equivalent to
those obtained purely using the metric formulation,
up to a \bdt/ in the presymplectic form. 

After setting up some preliminary notation and results 
in \secref{preliminaries},
we will first consider a Dirichlet \bdc/
as explained in \secref{Dirichletbc}.
Then we will carry out details of the Noether charge analysis
with the Dirichlet \bdc/
using the tetrad formulation of General Relativity
in \secref{Dbcanalysis}. 
The resulting covariant Dirichlet Hamiltonian for General Relativity 
is summarized in \secref{Dhamiltonian}
where we will discuss the equivalence between 
the metric and tetrad formulations. 
In \secref{generalbc} we will investigate 
a Neumann \bdc/ and corresponding Hamiltonian, 
along with more general \bdc/s and Hamiltonians. 
The main result will be to establish uniqueness of
mixed Dirichlet-Neumann \bdc/s for existence of a
Hamiltonian formulation of General Relativity. 
Finally, in \secref{covfieldeqs}
we will briefly discuss the form of the Dirichlet and Neumann 
covariant Hamiltonians,
and relate these to an analysis of boundary terms 
for the ADM Hamiltonian using the standard (non-covariant) 
ADM canonical variables.

\subsection{Preliminaries}
\label{preliminaries}

On a given smooth orientable 4-dimensional spacetime manifold $M$, 
let $\g{ab}{}$ be the spacetime metric tensor, 
$\vol{abcd}{}(g)$ be the volume form 
normalized with respect to the metric, 
and $\covgder{a}$ be the covariant (torsion-free) derivative operator 
determined by the metric. 

Now, let $\tfvec{a}$ be a complete, smooth timelike vector field on $M$,
and let $\Sigma$ be a region contained in a spacelike hypersurface
with the boundary of the region being 
a closed orientable 2-surface $\partial\Sigma$.
Let $\s{a}{}$ denote the unit outward spacelike normal to $\partial\Sigma$
orthogonal to $\tfvec{a}$,
let $\t{a}{}$ denote the unit future timelike normal to $\partial\Sigma$
orthogonal to $\s{a}{}$. 
Denote the metric tensor and volume form on $\partial\Sigma$ by 
\EQs
&& \metric{ab}{} =\g{ab}{} -\s{}{a}\s{}{b} +\t{}{a}\t{}{b} ,\\
&& \vol{ab}{} = \vol{abcd}{}(g) \s{c}{} \t{d}{} . 
\endEQs
This yields the decompositions
\EQs
&& \g{ab}{} = \metric{ab}{} +\s{}{a}\s{}{b} -\t{}{a}\t{}{b} , 
\label{metricdecomp}\\
&& \vol{abcd}{}(g) = 12\t{}{[a} \s{}{b} \vol{cd]}{} . 
\label{voldecomp}
\endEQs
Note that one has 
\EQ\label{orthog}
\tfvec{a} =\N{}{} \t{a}{} +\N{a}{} ,\quad 
\N{a}{} \s{}{a} = \tfvec{a} \s{}{a} =0
\endEQ
for some scalar function $\N{}{}$ and vector function $\N{a}{}$
on $\partial\Sigma$. 
It is convenient to extend the previous structures 
off $\partial\Sigma$ as follows. 
Let $\region$ be the spacetime region foliated by the images of $\Sigma$
under a one-parameter diffeomorphism on $M$ generated by $\tfvec{a}$,
and let $\bcsurface$ be the timelike boundary of $\region$ foliated by 
the images of $\partial\Sigma$.
Fix a time function $t$ which is constant 
on each of the spacelike slices 
diffeomorphic to $\Sigma$ under $\tfvec{a}$ in $\region$ 
and which is normalized by $\tfvec{a}\der{a} t=1$,
such that $t=0$ corresponds to $\Sigma$. 
Then $\bcsurface$ is a timelike hypersurface in $M$ 
whose intersection with spacelike hypersurfaces 
$\Sigma_t$ given by $t=\const$ in $\region$
consists of spacelike 2-surfaces $\partial\Sigma_t$
diffeomorphic to $\partial\Sigma$.
Finally, let 
$\s{a}{}$, $\t{a}{}$, $\metric{ab}{}$, $\vol{ab}{}$, $\N{}{}$, $\N{a}{}$
be extended to $\partial\Sigma_t$,
and let $\n{}{a}$ denote the unit future timelike normal to $\Sigma_t$
parallel to $\der{a} t$.

Note that, by construction, 
$\s{}{a}$ is hypersurface orthogonal to $\bcsurface$
and hence
\EQ\label{sorthog}
\s{}{[c} \der{b} \s{}{a]} =0 . 
\endEQ
If $\t{}{a}$ is expressed as a linear combination of $\s{}{a},\der{a} t$,
then since $\der{a} t$ obviously is 
hypersurface orthogonal to $\partial\Sigma_t$, 
it follows that 
\EQ\label{torthog}
\s{}{[d} \t{}{c} \der{b} \t{}{a]} =0 . 
\endEQ
In addition, note that 
$\s{a}{} \der{a} t$ measures the extent to which 
$\Sigma_t$ fails to be orthogonal to $\bcsurface$. 

Let $\Prop{\partial\Sigma}{a}{b}$ and $\Prop{t}{a}{b}$ 
be coordinate projection operators onto the respective tangent spaces of 
the 2-surface $\partial\Sigma_t$ 
and the integral curve of $\t{a}{}$,
and let 
$\Prop{\bcsurface}{a}{b}
= \Prop{\partial\Sigma}{a}{b} + \Prop{t}{a}{b}$, 
which is the projection operator onto the tangent space of
the timelike hypersurface $\bcsurface$.
Note that these operators are independent of the spacetime metric,
as they involve only the manifold structure of 
$\bcsurface$ and $\partial\Sigma_t$ in local coordinates in $M$.

Hereafter we work in terms of an \onfr/ $\frame{\mu}{a}$ (\ie/ tetrad)
for $\g{ab}{}$ on $M$.
The frame components of 
$\s{a}{}$, $\t{a}{}$, $\n{a}{}$, $\metric{ab}{}$, $\vol{ab}{}$,
$\g{ab}{}$, $\vol{abcd}{}(g)$
are given by 
\EQs
&& 
\s{\mu}{} =\s{a}{}\frame{\mu}{a} ,\
\t{\mu}{} =\t{a}{}\frame{\mu}{a} ,\
\n{\mu}{} =\n{a}{}\frame{\mu}{a} ,\
\label{frcoeff}\\
&& 
\metric{}{\mu\nu} = \metric{}{ab} \frame{\mu}{a}\frame{\nu}{b} ,\
\vol{}{\mu\nu} = \vol{}{ab} \frame{\mu}{a}\frame{\nu}{b} ,\\
&&
\flat{}{\mu\nu} = \g{}{ab} \frame{\mu}{a}\frame{\nu}{b} ,\
\vol{}{\alpha\beta\mu\nu} = \vol{}{abcd}(g) 
\frame{\alpha}{a}\frame{\beta}{b}\frame{\mu}{c}\frame{\nu}{c} ,
\endEQs
where 
$\flat{}{\mu\nu}=\metric{}{\mu\nu}+\s{\mu}{}\s{\nu}{}-\t{\mu}{}\t{\nu}{}
= diag(-1,1,1,1)$ is the Minkowski frame-metric,
with $\metric{}{\mu\nu}= diag(0,0,1,1)$. 
This leads to an \onfr/ for the metric $\metric{ab}{}$, 
given by 
\EQ
\metric{a}{\mu} 
= \metric{a}{b} \frame{\mu}{b}
\endEQ
satisfying
\EQ
\s{}{\mu} \metric{a}{\mu} = \t{}{\mu} \metric{a}{\mu} = 0 . 
\endEQ
Let the inverse \onfr/ for $\g{ab}{}$ and for $\metric{ab}{}$
be denoted by 
\EQ
\frame{a}{\mu} = \g{}{ab}\frame{\mu}{b} ,\
\metric{\mu}{a} = \metric{}{ab}\metric{b}{\mu} . 
\endEQ
Then, one has the decompositions
\EQ\label{frdecomp}
\frame{\mu}{a} =\metric{a}{\mu} +\s{}{a}\s{\mu}{}-\t{}{a}\t{\mu}{} ,\quad
\frame{a}{\mu} =\metric{\mu}{a} +\s{a}{}\s{}{\mu}-\t{a}{}\t{}{\mu} .
\endEQ

For later use, we will partially fix the SO(3,1) local gauge freedom 
in $\frame{a}{\mu}$ by choosing the coefficients $\s{}{\mu},\t{}{\mu}$
in the frame decomposition \eqref{frdecomp}
to be fixed functions on $M$, 
so that under a variation $\delta\g{ab}{}$,
\EQ\label{gaugefr}
\delta\s{}{\mu} =\delta\t{}{\mu} =\delta\metric{\mu\nu}{} =0
\endEQ
and hence, correspondingly, 
\EQs
&& 
\delta\s{}{a} =\s{}{\mu}\delta\frame{\mu}{a} ,\
\delta\t{}{a} =\t{}{\mu}\delta\frame{\mu}{a} , 
\\&& 
\delta\metric{ab}{} 
=2\metric{\mu\nu}{}\frame{\mu}{(a}\delta\frame{\nu}{b)}
=2\metric{(a}{\mu}\delta\metric{b)\mu}{} . 
\label{moregaugefr}
\endEQs
Similarly, one then also has
\EQs
&&
\delta\vol{abcd}{}(\theta) 
= 4\vol{\alpha\beta\mu\nu}{} 
\frame{\alpha}{[a} \frame{\beta}{b} \frame{\mu}{c} \delta\frame{\nu}{d]}
= \vol{abcd}{}(\theta) \frame{c}{\mu} \delta\frame{\mu}{c} , 
\\&&
\delta\vol{ab}{} 
= 2\vol{\mu\nu}{}\frame{\mu}{[a}\delta\frame{\nu}{b]}
= \vol{ab}{} \metric{\mu}{c}\delta\metric{c}{\mu} , 
\label{varvolmetric}
\endEQs
and thus
\EQ
\delta\vol{\mu\nu}{} = \delta\vol{\alpha\beta\mu\nu}{} =0 . 
\endEQ 
Consequently, some useful identities are given by
\EQs
&& 
\delta\frame{\mu}{a} 
= \delta\metric{a}{\mu} +\s{\mu}{} \delta\s{}{a} -\t{\mu}{} \delta\t{}{a} , 
\\&& 
\delta\frame{a}{\mu} 
= -\frame{a}{\nu} \frame{c}{\mu}\delta\frame{\nu}{c}
= \delta\metric{\mu}{a} +\s{}{\mu}{} \delta\s{a}{} -\t{}{\mu} \delta\t{a}{} . 
\endEQs

Now, a variation of the spacetime metric $\delta\g{ab}{}$
can be decomposed into the parts
\EQ
\delta\g{ab}{} = 
\delta\metric{ab}{} +2\s{}{(a}\delta\s{}{b)} -2\t{}{(a}\delta\t{}{b)} . 
\endEQ
By hypersurface orthogonality, one has the identities
\EQ\label{varst}
\delta\s{}{a} 
= \s{}{a} \s{b}{}\delta\s{}{b} ,\quad
\delta\t{}{a} 
= \s{}{a} \s{b}{}\delta\t{}{b} -\t{}{a} \t{b}{}\delta\t{}{b}
\endEQ
and
\EQ
\delta\metric{ab}{} 
=\metric{a}{c}\metric{b}{d} \delta\metric{cd}{}
+\metric{a}{c}\s{}{b}\s{d}{} \delta\metric{cd}{}
-\metric{a}{c}\t{}{b}\t{d}{} \delta\metric{cd}{} . 
\endEQ
Then, from the relation 
\EQ
\delta\g{}{ab} = 
\delta\metric{}{ab}{} +2\s{(a}{}\delta\s{b)}{} -2\t{(a}{}\delta\t{b)}{}
=-\g{}{ac}\g{}{bd}\delta\g{cd}{} , 
\endEQ
it straightforwardly follows that
\EQs
&& \delta\metric{}{ab} 
= -\metric{}{ac}\metric{}{bd}\delta\metric{cd}{} ,  
\label{varmetric}\\
&& \delta\s{a}{}
=\metric{b}{a} \delta\s{b}{}+\s{a}{}\s{}{b}\delta\s{b}{}
-\t{a}{}\t{}{b}\delta\s{b}{} , 
\label{vars}\\
&& \delta\t{a}{}
=\metric{b}{a} \delta\t{b}{}-\t{a}{}\t{}{b}\delta\t{b}{} , 
\label{vart}
\endEQs
where
\EQs
&& 
\metric{ab}{} \delta\s{b}{} =-\s{b}{}\delta\metric{ab}{} ,\ 
\metric{ab}{} \delta\t{b}{} =-\t{b}{}\delta\metric{ab}{} ,\\
&& 
\s{}{b} \delta\s{b}{} =-\s{b}{}\delta\s{}{b} ,\
\t{}{b} \delta\s{b}{} =-\t{b}{}\delta\s{}{b} ,\
\t{}{b} \delta\t{b}{} =-\t{b}{}\delta\t{}{b} ,
\label{decompvarst}
\endEQs
and again by hypersurface orthogonality, 
\EQ
\s{}{a}\delta\metric{}{ab}{}=\t{}{a}\delta\metric{}{ab}{} =0 ,\quad
\s{}{a} \delta\t{a}{} =0 .
\endEQ

Thus, the linearly independent parts of $\delta\frame{\mu}{a}$,
or equivalently of $\delta\frame{a}{\mu}$, 
are given by 
\EQ
\delta\metric{\mu}{a},
\metric{b}{a} \delta\s{b}{} ,\
\metric{b}{a} \delta\t{b}{} ,\
\s{}{b} \delta\s{b}{} ,\
\t{}{b} \delta\s{b}{} ,\
\t{}{b} \delta\t{b}{} .
\endEQ
Throughout, the time-flow vector field $\tfvec{a}$ is taken to be fixed,
$\delta\tfvec{a}=0$, under variations of $\frame{\mu}{a}$.

\subsection{Dirichlet boundary condition}
\label{Dirichletbc}

There is a natural motivation for a Dirichlet \bdc/
on the gravitational field in the Einstein equations
in analogy with the Maxwell equations
where the tangential components of 
the electromagnetic field potential $\A{a}$
are specified at the boundary. 
For General Relativity, similarly, one can introduce 
a Dirichlet \bdc/ given by specifying the tangential components of 
the spacetime metric $\g{ab}{}$ at the 2-surfaces $\partial\Sigma_t$. 
This \bdc/ is expressed equivalently by conditions 
on the variation of the metric tensor
\EQ\label{Dbc} 
\delta\metric{ab}{} |_{\partial\Sigma_t} =0 ,\quad
\delta\t{}{a} |_{\partial\Sigma_t} =0 ,\quad
t\geq 0 . 
\endEQ
Geometrically, this means that the metric given by 
\EQ
\h{ab}{} =\metric{ab}{} -\t{}{a}\t{}{b}
\endEQ
on the timelike boundary hypersurface $\bcsurface$ is specified data,
so it is held fixed under variations of $\g{ab}{}$, 
\EQ\label{geometricDbc}
\delta\h{ab}{}=0 
\quad\eqtext{ on $\bcsurface$.}
\endEQ
The geometrical form \eqref{geometricDbc} of the Dirichlet \bdc/ 
is often introduced when one considers 
an action principle for General Relativity
on a spacetime manifold 
with a fixed global timelike boundary hypersurface
\cite{Wald-book,York,Brown-York1,Brown-York2}.
We will see in the next section that this boundary condition 
in the form \eqref{Dbc} emerges naturally from 
the Noether charge analysis for the existence of a Hamiltonian
formulation of General Relativity 
for a spatially bounded local spacetime region. 

Note that, from the relations \eqsref{varmetric}{decompvarst},
one can decompose the Dirichlet \bdc/ \eqref{Dbc} into
an intrinsic part 
\EQ\label{intrinsicDbc}
\delta\h{}{ab}|_{\partial\Sigma_t}
= -\h{}{ac}\h{}{bd}\delta\h{cd}{}|_{\partial\Sigma_t}
=0 ,\quad 
t\geq 0
\endEQ
and an extrinsic part
\EQ\label{extrinsicDbc}
\h{ab}{}\delta\s{b}{}|_{\partial\Sigma_t}
= - \s{b}{}\delta\h{ab}{}|_{\partial\Sigma_t}
=0 ,\quad 
t\geq 0
\endEQ
with respect to the timelike hypersurface $\bcsurface$. 
The intrinsic part corresponds to fixing just the metric 
$\Pr{\bcsurface} \h{ab}{}$
restricted to the tangent space of $\bcsurface$,
where the projection $\Pr{\bcsurface}$ 
removes components of the hypersurface metric 
proportional to $\s{}{a}$.
Correspondingly, note that the volume form 
\EQ
\vol{abc}{}(h) =\vol{abcd}{}(g) \s{d}{} =3\vol{[ab}{} \t{}{c]}
\endEQ
on this surface is also fixed, 
$\delta\vol{abc}{}(h)=0$ on $\bcsurface$, 
since
\EQ
\delta\vol{abc}{}(h)
= \frac{1}{2}\vol{abc}{}(h)\h{}{mn}\delta\h{mn}{}
= -\frac{1}{2}\vol{abc}{}(h)\h{mn}{}\delta\h{}{mn} . 
\endEQ

The Dirichlet \bdc/ has a simple formulation in terms of
the \onfr/ $\frame{\mu}{a}$. 
It is convenient to introduce a frame for the metric $\h{ab}{}$ by
\EQ\label{frhdecomp}
\h{a}{\mu} = \h{a}{b}\frame{\mu}{b} 
= \metric{a}{\mu} -\t{}{a}\t{\mu}{} , 
\endEQ
and inverse frame 
\EQ
\h{}{a\mu} = \h{}{ab} \h{b}{\mu} . 
\endEQ
Then the Dirichlet \bdc/ \eqref{Dbc} is equivalent to 
\EQ
\delta\h{a}{\mu} |_{\partial\Sigma_t} =0 ,\quad 
t\ge 0,
\endEQ
with intrinsic part
\EQ
\delta\h{}{a\mu} |_{\partial\Sigma_t} =0 ,\quad
t\ge 0,
\endEQ
which is equivalent to \Eqref{intrinsicDbc}. 
These equivalences are immediate consequences of the identities
$\h{ab}{} = \h{a}{\mu} \h{b}{\nu} \flat{\mu\nu}{}$
and $\h{}{ab} = \h{\mu}{a} \h{\nu}{b} \flat{}{\mu\nu}$.
From these identities, 
one also has 
\EQs
&&
\delta\h{ab}{} =2\h{(a}{\mu} \delta\h{b)\mu}{} ,\\
&&
\delta\vol{abc}{}(h) = \h{\mu}{d} \delta\h{d}{\mu} .
\label{varvolhfr}
\endEQs
An additional useful identity is given by 
\EQ\label{varhid}
\delta\h{a}{c}
= -( \s{c}{} \delta\s{}{a} + \s{}{a} \delta\s{c}{} )
= -\s{}{a} \h{b}{c} \delta\s{b}{}
\endEQ
and therefore $\Pr{\bcsurface} \delta\h{a}{c}=0$. 

Finally, 
note that the intrinsic part of the Dirichlet \bdc/ on the frame
decomposes into
\EQ\label{intrinsicDbcdecomp}
\delta\metric{}{a\mu} |_{\partial\Sigma_t} =0 ,\quad
\delta\t{a}{} |_{\partial\Sigma_t} =0 ,\quad
t\ge 0 . 
\endEQ
The full, extrinsic Dirichlet \bdc/ is necessary and sufficient
for $\delta\h{a}{c} |_{\partial\Sigma_t} =0, t\geq 0$.

\subsection{Noether charge analysis}
\label{Dbcanalysis}

We consider the standard tetrad formulation of General Relativity,  
using an \onfr/ $\frame{\mu}{a}$ for $\g{ab}{}$
and a frame-connection 
\EQ\label{frcon}
\frcon{a}{\mu\nu}(\theta) = \frame{b\mu}{} \covgder{a}\frame{\nu}{b}
= 2\frame{b[\mu}{} \der{[a} \frame{\nu]}{b]} 
- \frame{b\mu}{}\frame{c\nu}{} \frame{}{a\alpha}\der{[b}\frame{\alpha}{c]} . 
\endEQ
Here the expression in the second equality is obtained from 
the relation 
\EQ\label{frconeq}
\frame{\nu}{[b} \frcon{a]\nu}{\mu}(\theta)
= \covgder{[a}\frame{\nu}{b]}
= \der{[a} \frame{\mu}{b]} . 
\endEQ
The curvature of this connection \eqref{frcon} is given by 
\EQ\label{frcurv}
\curv{ab}{\mu\nu}(\theta) 
= 2\der{[a}\frcon{b]}{\mu\nu}(\theta)
+ 2\frcon{[a}{\mu\sigma}(\theta)\frcon{b]\sigma}{\nu}(\theta)
= \curv{abcd}{}(g) \frame{c\mu}{}\frame{d\nu}{} , 
\endEQ
related to the Riemann curvature tensor $\curv{abcd}{}(g)$ of $\g{ab}{}$.

With $\frame{\mu}{a}$ as the field variable, 
the Lagrangian 4-form for General Relativity 
(without matter sources) is given by 
\EQ\label{Lfr}
\L{abcd}(\theta) = \vol{abcd}{}(\theta)\scurv(\theta) 
= 6\frame{\mu}{[a}\frame{\nu}{b}\tcurv{cd]\mu\nu}{}(\theta)
\endEQ
where 
\EQ\label{dualcurvid}
\tcurv{cd}{\mu\nu}(\theta) 
= \curv{cd\alpha\beta}{}(\theta) \vol{}{\alpha\beta\mu\nu}
=  2\der{[a}\tfrcon{b]}{\mu\nu}
-\frcon{[a}{\sigma[\mu}(\theta) \tfrcon{b]\sigma}{\nu]}(\theta) 
\endEQ
in terms of 
$\tfrcon{a\mu\nu}{}(\theta) 
=  \frcon{a}{\alpha\beta}(\theta) \vol{\alpha\beta\mu\nu}{}$. 
Then the variation of $\L{abcd}(\theta)$ gives, 
after integration by parts and use of the connection equation \eqref{frconeq},
\EQs
\frac{1}{6} \delta\L{abcd}(\theta)
&&
= 2 \delta\frame{\mu}{[a} ( \frame{\nu}{b}\tcurv{cd]\mu\nu}{}(\theta) )
+\frac{1}{6} \der{[a} \T{bcd]}(\theta,\delta\theta)
\label{varLtheta}
\endEQs
where 
\EQ\label{Tfr}
\T{bcd}(\theta,\delta\theta) 
= 12 \frame{\mu}{[c} \frame{\nu}{d} \delta\tfrcon{b]\mu\nu}{}(\theta) 
=8\vol{ebcd}{} \frame{\alpha}{e} \frame{a}{\beta} 
\delta\frcon{a}{\alpha\beta}(\theta)
\endEQ
defines the symplectic potential 3-form. 
The field equations for $\frame{\mu}{a}$, 
obtained from the coefficient of $\delta\frame{\mu}{a}$ in \Eqref{varLtheta},
are given by 
\EQ\label{frfieldeq}
\E\mixedindices{\mu}{bcd}(\theta)
= 12\frame{\nu}{[b} \tcurv{cd]\mu\nu}{}(\theta)
= 8 \vol{bcda}{}(\theta) ( 
\curv{}{a\mu}(\theta) -\frac{1}{2}\frame{a\mu}{}\scurv(\theta) )
=0 . 
\endEQ
Thus $\frame{\mu}{a}$ satisfies 
$\curv{}{a\mu}(\theta)=0$, 
which is equivalent to the vacuum Einstein equations for the spacetime metric
\EQ\label{GReq}
\curv{ab}{}(g)=0
\endEQ 
arising as the stationary points of the action functional 
$S(g) = \int_M \vol{abcd}{}(g)\scurv(g)$
under compact support variations of $\g{ab}{}$. 

The Noether current associated to $\tfvec{a}$ is given by 
the 3-form
\EQs
\J{abc}(\xi,\theta)
&& 
= \T{abc}(\theta,\Lie{\xi}\theta) +4\tfvec{d}\L{abcd}(\theta)
\nonumber\\&&
= 12\frame{\mu}{[b} \frame{\nu}{c} \Lie{\xi}\tfrcon{a]\mu\nu}{}(\theta)
+24\tfvec{d} \frame{\mu}{[c} \frame{\nu}{d} \tcurv{ab]\mu\nu}{}(\theta)
\label{Jfr}
\endEQs
with the first term obtained from 
the variation of the frame connection \eqref{frcon} 
after replacement of $\delta\frame{\mu}{a}$ by 
the Lie derivative
$\Lie{\xi} \frame{\mu}{a} 
=\tfvec{e} \der{e} \frame{\mu}{a} +\frame{\mu}{e} \der{a}\tfvec{e}$
and use of the fact that Lie derivatives commute with 
exterior (skew) derivatives. 
We now simplify the first term in \Eqref{Jfr} as follows. 
First we express
\EQs
\Lie{\xi}\frcon{a}{\mu\nu}(\theta)
&&
= \der{a}( \tfvec{e}\frcon{e}{\mu\nu}(\theta) )
+ \tfvec{e}( \curv{ea}{\mu\nu}(\theta) 
- 2\frcon{[e}{\mu\sigma}(\theta) \frcon{a]\sigma}{\nu}(\theta) ) . 
\endEQs
Hence we obtain
\EQ\label{dualfrconlieid}
12 \frame{\mu}{[b} \frame{\nu}{c} \Lie{\xi}\tfrcon{a]\mu\nu}{}(\theta)
= \der{[a} ( 12 \frame{\mu}{b} \frame{\nu}{c]} 
\tfvec{e}\tfrcon{e\mu\nu}{}(\theta) )
-12 \tfvec{e} \frame{\mu}{[b} \frame{\nu}{c} \tcurv{a]e\mu\nu}{}(\theta) 
\endEQ
through use of the identity \eqref{dualcurvid}.
Next we combine the second terms in both \Eqrefs{Jfr}{dualfrconlieid}
to get
\EQ
-12 \tfvec{e} \frame{\mu}{[b} \frame{\nu}{c} \tcurv{a]e\mu\nu}{}(\theta) 
+24\tfvec{d} \frame{\mu}{[c} \frame{\nu}{d} \tcurv{ab]\mu\nu}{}(\theta)
= \tfvec{e} \vol{abcd}{}(g) ( 4\id{e}{d} \scurv(g) -8\curv{e}{d}(g) ) . 
\endEQ
Thus, one obtains the Noether current
\EQ\label{simpleJfr}
\J{abc}(\xi;\theta)
= 3\der{[a} \Q{bc]}(\xi;\theta) 
-\tfvec{e} \frame{}{e\mu} \E\mixedindices{\mu}{abc}(\theta)
\endEQ
where
\EQ\label{Qfr}
\Q{bc}(\xi;\theta) 
= 4\tfvec{d} \tfrcon{d\mu\nu}{}(\theta) \frame{\mu}{b} \frame{\nu}{c} 
\endEQ
is the Noether current potential 2-form. 

On vacuum solutions $\frame{\mu}{a}$, 
the Noether current reduces to an exact 3-form
\EQ
\J{abc}(\xi;\theta) 
= 3\der{[a} \Q{bc]}(\xi;\theta) . 
\endEQ
Therefore, the Noether charge for vacuum solutions 
is given by the boundary 2-surface integral 
\EQs
Q_\Sigma(\xi) 
= \int_\Sigma \J{abc}(\xi;\theta) &&
= \int_{\partial\Sigma} \vol{bc}{} 
4 \tfvec{d} \tfrcon{d\mu\nu}{}(\theta) \vol{}{\mu\nu}
= \int_{\partial\Sigma} 
8 \tfvec{d} \frcon{d\mu\nu}{}(\theta) \t{\mu}{} \s{\nu}{} dS
\label{Qfrsol}
\endEQs
where 
$dS$ is the volume element on $\partial\Sigma$ 
corresponding to the volume form $\vol{bc}{}$ in local coordinates. 

Now the symplectic current,
defined by the antisymmetrized variation of $\T{bcd}(\theta,\delta\theta)$,
is given by the 3-form
\EQ\label{wfr}
\frac{1}{24} \w{bcd}(\theta,\delta_1\theta,\delta_2\theta)
= \frame{\mu}{[c} \delta_1\frame{\nu}{d} \delta_2\tfrcon{b]\mu\nu}{}(\theta)
- \frame{\mu}{[c} \delta_2\frame{\nu}{d} \delta_1\tfrcon{b]\mu\nu}{}(\theta) . 
\endEQ
Then the presymplectic form on $\Sigma$ is defined by
\EQ
\Omega_\Sigma(\theta,\delta_1\theta,\delta_2\theta)
= \int_\Sigma \w{bcd}(\theta,\delta_1\theta,\delta_2\theta)
= 24 \int_\Sigma
\frame{\mu}{[c} \delta_1\frame{\nu}{d} \delta_2\tfrcon{b]\mu\nu}{}(\theta)
- \frame{\mu}{[c} \delta_2\frame{\nu}{d} \delta_1\tfrcon{b]\mu\nu}{}(\theta) .
\endEQ
A Hamiltonian conjugate to $\xi$ on $\Sigma$ 
is a function 
$H_\Sigma(\xi;\theta)=\int_\Sigma \Hdens\downindex{abc}(\xi;\theta)$
for some locally constructed 3-form $\Hdens\downindex{abc}(\xi;\theta)$
such that
\EQ\label{Heqfr}
\delta H_\Sigma(\xi;\theta) \equiv H'_\Sigma(\xi;\theta,\delta\theta)
= \Omega_\Sigma(\delta\theta, \Lie{\xi}\theta)
\endEQ
holds for arbitrary variations $\delta\frame{\mu}{a}$ 
away from vacuum solutions $\frame{\mu}{a}$.

In terms of the Noether current \eqref{simpleJfr}, 
the presymplectic form on $\Sigma$ yields
\EQs
\Omega_\Sigma(\theta,\delta\theta,\Lie{\xi}\theta)
&&
= \int_\Sigma \delta\J{abc}(\xi;\theta) 
-4 \tfvec{d} \E\mixedindices{\mu}{[abc}(\theta) \delta\frame{}{d]\mu} 
-\int_{\partial\Sigma} \tfvec{c}\T{abc}(\theta,\delta\theta) . 
\label{presymplecticfreq}
\endEQs
Consequently, for variations $\delta\frame{\mu}{a}$ 
with compact support on the interior of $\Sigma$, 
the Noether current defines a Hamiltonian conjugate to $\tfvec{a}$, 
\EQ
H(\xi;\theta) 
= \int_\Sigma \J{abc}(\xi;\theta) 
= 8 \int_\Sigma \tfvec{e} \frame{}{e\mu} 
\E\mixedindices{\mu}{abc}(\theta)
+ \int_{\partial\Sigma} \Q{ab}(\xi;\theta) , 
\endEQ
which is equal to the Noether charge \eqref{Qfrsol}
when $\frame{\mu}{a}$ is a vacuum solution. 
Explicitly, from \Eqrefs{frfieldeq}{Qfr}, 
one has
\EQs
H(\xi;\theta) 
&&
= 8 \int_\Sigma \vol{dabc}{} (
\tfvec{e} \frame{d}{\mu} \curv{e}{\mu}(\theta) 
-\frac{1}{2}\tfvec{d} \scurv(\theta) ) 
+ 4 \int_{\partial\Sigma} \vol{bc}{} 
\tfvec{d} \tfrcon{d\mu\nu}{}(\theta) \vol{}{\mu\nu}
\nonumber\\&&
= 8 \int_\Sigma \tfvec{e} \n{}{\mu}
( \curv{e}{\mu}(\theta) -\frac{1}{2}\frame{\mu}{e} \scurv(\theta) ) d\Sigma
+ 8\int_{\partial\Sigma} 
\tfvec{d} \frcon{d\mu\nu}{}(\theta) \t{\mu}{} \s{\nu}{} dS
\endEQs
where $d\Sigma$ is the coordinate volume element on $\Sigma$ 
obtained from the volume form $\vol{abcd}{} \n{d}{}$.

To define a Hamiltonian $H_\Sigma(\xi;\theta)$
for variations $\delta\frame{\mu}{a}$ without compact support, 
it follows that the term 
$\tfvec{c}  \T{abc}(\theta,\delta\theta)$ in \Eqref{presymplecticfreq}
needs to be a total variation at the boundary $\partial\Sigma$,
i.e. there must exist a locally constructed 3-form 
$\B{abc}(\theta)$ such that one has
\EQ\label{bteq}
\tfvec{c} \T{abc}(\theta,\delta\theta) |_{\partial\Sigma}
=( \tfvec{c} \delta\B{abc}(\theta) 
+\der{[a}\a{}{b]}(\xi;\theta,\delta\theta) )|_{\partial\Sigma}
\endEQ
where $\a{}{b}(\xi;\theta, \delta\theta)$ is a locally constructed 1-form.
This equation is equivalent to 
\EQ\label{GRbceq}
\vol{}{ab}\tfvec{c} 
\T{abc}(\theta,\delta\theta)|_{\partial\Sigma}
=\vol{}{ab}\tfvec{c} \delta\B{abc}(\theta,\delta\theta)|_{\partial\Sigma}
+ \metric{c}{d}\der{d} \ta{c}{}(\xi;\theta,\delta\theta)|_{\partial\Sigma}
\endEQ
where 
$\ta{c}{}(\xi;\theta,\delta\theta)
= \vol{}{cb} \a{}{b}(\xi;\theta,\delta\theta)$
and the symplectic potential term is given by, 
using identity \eqref{voldecomp}, 
\EQs
\vol{}{ab}\tfvec{c} \T{abc}(\theta,\delta\theta)
&&
= 32 \tfvec{c} \t{}{[\alpha} \s{}{\nu]} 
\frame{\alpha}{c} \frame{a}{\mu} \delta\frcon{a}{\mu\nu}(\theta) . 
\label{simpleTfr}
\endEQs
Hence we now have the following result. 

\Proclaim{ Proposition 3.1. }{
A Hamiltonian conjugate to $\tfvec{a}$ on $\Sigma$ 
exists for variations $\delta\frame{\mu}{a}$
with support on $\partial\Sigma$ if and only if 
\EQ\label{frbcdeteq}
\vol{}{ab} \tfvec{c} \delta\B{abc}(\theta) |_{\partial\Sigma}
= 32 ( \tfvec{c} \t{}{[\alpha} \s{}{\nu]} \frame{\alpha}{c} \frame{a}{\mu}
\delta\frcon{a}{\mu\nu}(\theta) )|_{\partial\Sigma}
- \metric{c}{d}\der{d} \ta{c}{}(\xi;\theta,\delta\theta) |_{\partial\Sigma}
\endEQ
for some locally constructed 3-form 
$\B{abc}(\theta)$ in $T^*(\Sigma)$
and locally constructed vector
$\ta{c}{}(\xi;\theta,\delta\theta)$ in $T(\partial\Sigma)$. 
The Hamiltonian is given by 
the Noether charge plus an additional \bdt/
\EQs
H_\Sigma(\xi;\theta) 
&&
= \int_\Sigma \J{abc}(\xi;\theta) 
- \int_{\partial\Sigma} \tfvec{c}\B{abc}(\theta)
\nonumber\\&&
= 8 \int_\Sigma \tfvec{e} \frame{}{e\mu} 
\E\mixedindices{\mu}{abc}(\theta)
+ \int_{\partial\Sigma} \Q{ab}(\xi;\theta) - \tfvec{c}\B{abc}(\theta)
\nonumber\\&&
= 8 \int_\Sigma \tfvec{e} \n{}{\mu}
( \curv{e}{\mu}(\theta) -\frac{1}{2}\frame{\mu}{e} \scurv(\theta) ) d\Sigma
+ \int_{\partial\Sigma} \tfvec{d} (
8 \frcon{d\mu\nu}{}(\theta) \t{\mu}{} \s{\nu}{} -\frac{1}{2}\B{d}(\theta) ) dS
\label{defHfr}
\endEQs
with $\B{d}(\theta)= \vol{}{bc} \B{bcd}(\theta)$. }

We now show that the equation \eqref{frbcdeteq} 
for existence of a Hamiltonian \eqref{defHfr}
is satisfied for
the intrinsic Dirichlet \bdc/ on $\frame{\mu}{a}$, 
\EQ\label{intrinsicDbcfr}
\delta \h{}{a\mu} |_{\partial\Sigma_t} =0 ,\quad 
t\ge 0 , 
\endEQ
and then we derive the corresponding Hamiltonian \bdt/. 
Henceforth we take $\frame{\mu}{a}$ to satisfy 
the gauge conditions \eqsref{gaugefr}{moregaugefr}
naturally associated to the boundary hypersurface $\bcsurface$. 

Consider the left-side of \Eqref{frbcdeteq}.
The \bdc/ \eqref{intrinsicDbcfr} yields 
$\delta\metric{}{a\mu} |_{\partial\Sigma} =0$
and hence, from \Eqref{varvolmetric}, 
$\delta \vol{}{ab} 
= \vol{}{ab} \metric{c\mu}{} \delta\metric{}{c\mu} |_{\partial\Sigma} =0$.
Thus, we have
\EQ
\vol{}{ab} \tfvec{c} \delta\B{abc}(\theta) |_{\partial\Sigma}
= \delta( \vol{}{ab} \tfvec{c} \B{abc}(\theta) )  |_{\partial\Sigma} . 
\endEQ
Next consider the right-side of \Eqref{frbcdeteq}.
We integrate by parts with respect to the variation in the first term 
to get 
\EQ
\tfvec{c} \t{}{[\alpha} \s{}{\nu]} \frame{\alpha}{c} \frame{a}{\mu}
\delta\frcon{a}{\mu\nu}(\theta)
= \delta( 
\tfvec{c} \t{}{[\alpha} \s{}{\nu]} \frame{\alpha}{c} \frame{a}{\mu}
\frcon{a}{\mu\nu}(\theta) )
-\tfvec{c} \t{}{[\alpha} \s{}{\nu]} \frcon{a}{\mu\nu}(\theta)
( \frame{a}{\mu} \delta\frame{\alpha}{c} 
+ \frame{\alpha}{c} \delta\frame{a}{\mu} ) . 
\label{deteqterm}
\endEQ
Then, using orthogonality relations \eqrefs{orthog}{varst}, 
we find that the second term in \Eqref{deteqterm} vanishes as follows.
First,
\EQ
\tfvec{c} \t{}{[\alpha} \s{}{\nu]} \delta\frame{\alpha}{c} 
= \frac{1}{2} \tfvec{c}  ( \s{}{\nu} \delta\t{}{c} - \t{}{\nu} \delta\s{}{c} )
= -\s{}{\nu} \tfvec{c} \t{}{c} \t{}{b} \delta\t{b}{}
=0
\endEQ
since $\delta\t{b}{} |_{\partial\Sigma}=0$
by the \bdc/ \eqref{intrinsicDbcfr}. 
In addition,
\EQ
\tfvec{c} \t{}{[\alpha} \s{}{\nu]} \frame{\alpha}{c} 
\delta\frame{a}{\mu}
=  \frac{1}{2} \tfvec{c} \t{}{c} \s{}{\nu}
( \delta\h{\mu}{a} +\s{}{\mu}\delta\s{a}{} )
=  -\frac{1}{2} \N{}{} \s{}{\nu} \s{}{\mu} \delta\s{a}{} . 
\endEQ
Hence, the second term in \Eqref{deteqterm} 
reduces to 
\EQ
\frac{1}{2} \N{}{} \s{}{\nu} \s{}{\mu} \frcon{a}{\mu\nu}(\theta)
\delta\s{a}{}
=0
\endEQ
since $\frcon{a}{(\mu\nu)}(\theta) =0$. 

Consequently, returning to equation \eqref{frbcdeteq}, we obtain
\EQ
\delta( \tfvec{c} \B{c}(\theta) 
- 32 \t{}{[\alpha} \s{}{\nu]} \tfvec{c} \frame{\alpha}{c} \frame{a}{\mu}
\frcon{a}{\mu\nu}(\theta) ) |_{\partial\Sigma}
= - \metric{c}{d}\der{d} \ta{c}{}(\xi;\theta,\delta\theta) 
|_{\partial\Sigma}
\endEQ
which obviously is satisfied by 
\EQ\label{BDbcfr}
\tfvec{c} \B{c}(\theta) 
= 32 \t{}{[\alpha} \s{}{\nu]} 
\tfvec{c} \frame{\alpha}{c} \frame{d}{\mu} \frcon{d}{\mu\nu}(\theta)
\endEQ
and $\ta{c}{}(\xi;\theta,\delta\theta) =0$. 
This verifies Proposition~3.1
using the intrinsic Dirichlet \bdc/ \eqref{intrinsicDbcfr}. 

Finally, from expressions 
\eqref{BDbcfr} for $\tfvec{c} \B{c}(\theta)$
and \eqref{Qfr} for $\Q{bc}(\xi,\theta)$, 
we obtain a Hamiltonian \eqref{defHfr}
with the \bdt/ given by 
\EQs
H_B(\xi,\theta)
&& 
= \int_{\partial\Sigma} 
\Q{ab}(\xi,\theta) -\tfvec{c}\B{abc}(\theta,\delta\theta)
\nonumber\\&&
= 8 \int_{\partial\Sigma} \vol{ab}{} \tfvec{c}
( \t{}{\mu} \s{}{\nu} \frcon{c}{\mu\nu}(\theta)
-2  \t{}{[\alpha} \s{}{\nu]} 
\frame{\alpha}{c} \frame{d}{\mu} \frcon{d}{\mu\nu}(\theta) ) . 
\endEQs
Hence, the Hamiltonian \bdt/ takes the form
\EQ\label{Hfr}
H_B(\xi,\theta)
= 8 \int_{\partial\Sigma} \tfvec{c} \P{}{c}(\theta) dS
\endEQ
where
$\P{}{c}(\theta)
= \t{}{\mu} \s{}{\nu} \frcon{c}{\mu\nu}(\theta) 
- \t{}{c} \s{}{\nu} \frame{a}{\mu} \frcon{a}{\mu\nu}(\theta)
+ \s{}{c} \t{}{\nu} \frame{a}{\mu} \frcon{a}{\mu\nu}(\theta)$. 
This expression is simplified 
by the identities \eqrefs{metricdecomp}{frdecomp},
which yield
\EQ\label{PDfr}
\P{}{c}(\theta)
= \t{a}{} \s{}{\nu} \metric{c}{d}\covgder{d} \frame{\nu}{a}
-\t{}{c} \s{}{\nu} \metric{}{bd}\covgder{b} \frame{\nu}{d}
+ \s{}{c} \t{}{\nu} \metric{}{bd}\covgder{b} \frame{\nu}{d} . 
\endEQ
Thus, we have the following main result. 

\Proclaim{ Theorem 3.2. }{
For the intrinsic Dirichlet \bdc/ \eqref{intrinsicDbcfr},
a Hamiltonian conjugate to $\tfvec{a}$ on $\Sigma$ is given by 
\EQ\label{fullDHfr}
H_\Sigma(\xi;\theta)
= 8 \int_\Sigma \tfvec{e} \n{}{\mu}
( \curv{e}{\mu}(\theta) -\frac{1}{2}\frame{\mu}{e} \scurv(\theta) ) d\Sigma
+ 8 \int_{\partial\Sigma} \tfvec{c} \P{}{c}(\theta) dS .
\endEQ
On vacuum solutions $\frame{\mu}{a}$, 
the Hamiltonian reduces to the surface integral \eqref{Hfr}, \eqref{PDfr}. 
}

Note, this Hamiltonian is unique up to adding 
an arbitrary covector function of the Dirichlet boundary data $\h{\mu}{a}$
to $\P{}{c}(\theta)$.

\subsection{Dirichlet Hamiltonian}
\label{Dhamiltonian}

On vacuum solutions of the Einstein equations, 
the Hamiltonian \eqref{fullDHfr}
with the Dirichlet \bdc/ \eqref{intrinsicDbcfr} 
holding on the timelike hypersurface $\bcsurface$
bounding a local spacetime region $\region$
takes a simple form if the frame $\frame{\mu}{a}$ is adapted
to the boundary 2-surfaces $\partial\Sigma_t$ and $\tfvec{a}$. 
Let
\EQ\label{adaptedfr}
\adfr{0}{a} =\t{}{a}, 
\adfr{1}{a} =\s{}{a},
\adfr{2}{[a} \adfr{3}{b]} =\vol{ab}{},
\adfr{2}{a} =\vol{a}{b} \adfr{3}{b} 
\endEQ
which defines a preferred \onfr/ $\adfr{\mu}{a}$. 
It follows that
\EQ
\t{\nu}{} =\t{a}{} \adfr{\nu}{a}=-\id{0}{\nu} ,\
\s{\nu}{} =\s{a}{} \adfr{\nu}{a}=\id{1}{\nu} . 
\endEQ
This choice of frame is unique up to rotations of 
$\adfr{2}{a},\adfr{3}{a}$. 

\Proclaim{ Theorem 3.3. }{
For the Dirichlet \bdc/ \eqref{intrinsicDbcfr}, 
the Hamiltonian \eqref{fullDHfr} conjugate to $\tfvec{a}$ on $\Sigma$, 
evaluated in the \onfr/ \eqref{adaptedfr} 
for vacuum solutions $\adfr{\mu}{a}$, 
is given by the surface integral
\EQ\label{HDfr}
\H{D}(\xi;\vartheta)
= 8\int_{\partial\Sigma} \tfvec{c} \P{\rm D}{c}(\vartheta) dS
\endEQ
where
\EQ\label{PD}
\P{\rm D}{c}(\vartheta)
= \t{a}{} \metric{c}{d}\covgder{d} \s{}{a}
- \t{}{c} \metric{}{bd}\covgder{b} \s{}{d}
+ \s{}{c} \metric{}{bd}\covgder{b} \t{}{d} . 
\endEQ
}
We refer to $\H{D}(\xi;\vartheta)$ 
as the {\it Dirichlet Hamiltonian} 
for the gravitational field in the local spacetime region $\region$, 
and $\P{\rm D}{c}(\vartheta)$ 
as the {\it Dirichlet symplectic vector} 
associated to the boundary 2-surfaces $\partial\Sigma_t$.
Note that, 
since $\tfvec{a}$ lies in $\bcsurface$, 
only the first two terms in $\P{\rm D}{c}(\vartheta)$
contribute to $\H{D}(\xi;\vartheta)$. 
The significance of the full expression for $\P{\rm D}{c}(\vartheta)$
and its resulting geometrical properties 
are discussed in \Ref{symplecticvectors}.

The general form of the Hamiltonian boundary term \eqrefs{Hfr}{PDfr} 
differs from the special form \eqrefs{HDfr}{PD} 
when evaluated in an \onfr/ not adapted 
to $\partial\Sigma_t$ and $\tfvec{a}$.
In particular, we obtain the relation 
\EQ\label{PDeq}
\P{}{c}(\theta)= \P{\rm D}{c}(\vartheta)
-\metric{c}{d} \t{\nu}{} \der{d} \s{}{\nu}
+ \t{}{c} \metric{}{b\nu}\der{b} \s{}{\nu}
- \s{}{c} \metric{}{b\nu}\der{b} \t{}{\nu}
\endEQ
and so the general form of the symplectic vector $\P{}{c}(\theta)$ 
differs from $\P{\rm D}{c}(\vartheta)$ by various gradient terms. 
These terms can be understood by considering 
a change of \onfr/ 
\EQ\label{changefr}
\frame{\mu}{a} \rightarrow \U{\mu}{\nu}\frame{\nu}{a}
\endEQ
where $\U{\mu}{\nu}$ is a SO(3,1) transformation
acting in the frame bundle of the spacetime $(M,g)$
at $\partial\Sigma_t$. 
Such transformations are defined by 
$\U{-1\mu}{\nu}= \U{\alpha}{\beta} \flat{\alpha\nu}{} \flat{}{\mu\beta}$
and
$\det(U) =1$, 
where $\U{-1\mu}{\nu}$ is the inverse of $\U{\mu}{\nu}$, 
given by
$\U{-1\mu}{\nu} \U{\nu}{\alpha} = \id{\alpha}{\mu} 
= \U{\mu}{\nu} \U{-1\nu}{\alpha}$.

The transformations \eqref{changefr} are a gauge symmetry
of the tetrad formulation for General Relativity. 
Under the change of \onfr/, 
one has
\EQ\label{changefrcon}
\frcon{a\mu}{\nu}(\theta) 
= \frame{b}{\mu} \covgder{a}\frame{\nu}{b}
\rightarrow 
\frcon{a\mu}{\nu}(U\theta) 
= \U{-1\alpha}{\mu} \frame{b}{\alpha} 
\covgder{a}( \U{\nu}{\beta} \frame{\beta}{b} )
= \U{-1\alpha}{\mu} \U{\nu}{\beta} \frcon{a\alpha}{\beta}(\theta)
+ \U{-1\alpha}{\mu} \der{a}  \U{\nu}{\alpha} , 
\endEQ
and so, through substitution of the transformation \eqref{changefrcon}
into the curvature \eqref{frcurv}, 
\EQ
\curv{ab\mu}{\nu}(\theta) 
\rightarrow 
\curv{ab\mu}{\nu}(U\theta) 
= \U{-1\alpha}{\mu} \U{\nu}{\beta} \curv{ab\alpha}{\beta}(\theta)
\endEQ
after cancellations of terms. 
Hence the Lagrangian \eqref{Lfr} 
for the field variable $\frame{\mu}{a}$ 
is gauge invariant. 
As a consequence, it is straightforward to see that 
the symplectic structure given by 
the symplectic potential \eqref{Tfr} and current \eqref{wfr}
must be gauge invariant. 
In particular, 
note that one has
$\delta\frcon{a\mu}{\nu}(U\theta) 
= \U{-1\alpha}{\mu} \U{\nu}{\beta} \delta\frcon{a\alpha}{\beta}(\theta)$
where the gradient term from \Eqref{changefrcon}
drops out of the variation since it has no dependence on $\frame{\mu}{a}$.
This explicitly establishes the gauge invariance of 
$\T{abc}(\theta,\delta\theta)$ 
and hence of 
$\w{abc}(\theta,\delta_1\theta,\delta_2\theta)$. 

However, the Noether charge \eqref{Qfr} fails to be gauge invariant
due to its explicit dependence on the frame connection. 
Consequently, 
it follows that the gradient terms in the symplectic vector \eqref{PDeq}
originate from a gauge transformation on the frame connection 
under \eqref{changefr} as given by a transformation
relating the adapted \onfr/ to a general \onfr/, 
$\frame{\mu}{a} = 
\U{\mu}{0} \t{}{a} + \U{\mu}{1} \metric{a}{1} + \U{\mu}{2} \metric{a}{2}
+ \U{\mu}{4} \s{}{a}$. 

The gauge invariance of the symplectic structure
arising from the tetrad formulation of the Lagrangian
means that the symplectic potential \eqref{Tfr} and current \eqref{wfr} 
are equivalent to manifestly gauge-invariant expressions
derived using the metric formulation of General Relativity
with $\g{ab}{}$ as the field variable. 
It can be shown that one has
\EQ
\T{abc}(g,\delta g)
= \T{abc}(\theta,\delta\theta) 
+ 6\der{[c} \btT{ab]}(g,\delta g)
\endEQ
where $\btT{ab}(g,\delta g)$ is a locally constructed 2-form,
and so the symplectic potentials are equivalent to within an exact 3-form. 
This contributes a \bdt/ to the presymplectic form
obtained from the metric Lagrangian $\L{abcd}(g)= \vol{abcd}{}(g)\scurv(g)$, 
\EQ\label{gbt}
\Omega_\Sigma(g,\delta g,\Lie{\xi}g)
= \Omega_\Sigma(\theta,\delta\theta,\Lie{\xi}\theta)
+ \int_{\partial\Sigma} \vol{}{ab} 
( \delta\btT{ab}(g,\Lie{\xi}g) - \Lie{\xi} \btT{ab}(g,\delta g) ) dS . 
\endEQ
Correspondingly, 
the Noether charge 2-form $\Q{ab}(\xi,g)$
arising in the metric formulation
differs from $\Q{ab}(\xi,\theta)$ 
in the tetrad formulation 
by the term $2\btT{ab}(g,\Lie{\xi}g)$. 
Explicitly, using the metric Lagrangian, 
one finds that \cite{Wald-Iyer2}
\EQs
\frac{1}{4} \vol{}{ab} \Q{ab}(\xi,g)
&&
= -4 \t{}{[c} \s{}{d]} \covgcoder{c} \tfvec{d}
= 4 \tfvec{d} \t{c}{} \covgder{c} \s{}{d}
-2( \s{c}{} \Lie{\xi}\t{}{c} + \t{c}{} \Lie{\xi}\s{}{c} ) . 
\label{Qfreq}
\endEQs
Here the first term in \Eqref{Qfreq} is simply
the Noether charge \eqref{Qfr} evaluated in 
the adapted \onfr/ \eqref{adaptedfr}, 
\EQs
\vol{}{ab} \Q{ab}(\xi,\vartheta) 
&&
= 4 \tfvec{c} \frcon{c}{\mu\nu}(\vartheta) 
\vol{}{ab} \vol{abmn}{}(g) \adfr{m}{\mu} \adfr{n}{\nu} 
= 16 \tfvec{c} \t{}{\mu} \s{}{\nu} \adfr{b\mu}{} \covgder{c} \adfr{\nu}{b} 
= 16 \tfvec{c} \t{d}{} \covgder{c} \s{}{d}
\endEQs
since $\Pr{\bcsurface} (\covgder{[c} \s{}{d]})=0$
by hypersurface orthogonality of $\s{}{d}$. 
The second term in \Eqref{Qfreq} simplifies through
the hypersurface orthogonality relations \eqrefs{sorthog}{torthog},
leading to 
\EQ
\t{c}{} \Lie{\xi}\s{}{c}
= \t{c}{} ( 2 \tfvec{b} \covgder{[b} \s{}{c]}
+ \der{c}( \tfvec{b} \s{}{b} ) )
=0
\endEQ
and 
\EQ
\s{c}{} \Lie{\xi}\t{}{c}
= -\tfvec{a}{} \frac{\N{}{}}{\alpha} \der{a} \beta
\endEQ
where $\alpha,\beta,\N{}{}$ are scalar functions defined by 
\EQ
\s{}{a} =\alpha \der{a} s ,\quad
\t{}{a} =-\N{}{} ( \der{a} t + \beta \der{a} s) 
\endEQ
with 
\EQ
\Lie{\xi} t = \tfvec{e} \der{e} t =1 ,\quad
\Lie{\xi} s = \tfvec{e} \der{e} s =0 . 
\endEQ
Hence we obtain the relation
\EQ\label{compareQ}
\vol{}{ab} \Q{ab}(\xi,g)
= \vol{}{ab} \Q{ab}(\xi,\vartheta)
+ 8\tfvec{a}{} \frac{\N{}{}}{\alpha} \der{a} \beta . 
\endEQ

A similar relation can be shown to hold between
the respective symplectic vectors 
arising in the tetrad and metric Hamiltonian formulations
of General Relativity.  
In particular, by direct calculation with $\g{ab}{}$ as the field variable,
one finds that the full Dirichlet \bdc/ \eqref{Dbc}
yields a Hamiltonian conjugate to $\tfvec{a}$ on $\Sigma$
whose \bdt/ is given by 
\EQ\label{gDH}
\H{D}(\xi;g)
= 8 \int_{\partial\Sigma} \tfvec{c} \P{D}{c}(g) dS
\endEQ
where 
\EQ\label{gDP}
\P{\rm D}{c}(g)
= \P{\rm D}{c}(\vartheta) +\frac{\N{}{}}{\alpha} \der{c} \beta . 
\endEQ
This differs from the symplectic vector in the tetrad formulation
by the same gradient term
occurring in the Noether charges \eqref{compareQ}. 
The extrinsic part \eqref{extrinsicDbc} of
the Dirichlet \bdc/ is necessary in obtaining this Hamiltonian, 
because of the \bdt/ in the presymplectic form \eqref{gbt}. 
Interestingly, 
in the case when $\tfvec{a}$ is orthogonal to $\Sigma_t$, 
then $\beta=0$, 
and one finds that the weaker, 
intrinsic Dirichlet \bdc/ \eqref{intrinsicDbc}
is sufficient for existence of the metric Hamiltonian 
\eqrefs{gDH}{gDP}. 
Moreover, in this case 
the presymplectic form \eqref{gbt} 
and symplectic vector \eqref{gDP}
are exactly the same as those obtained 
in the tetrad formulation 
using the adapted \onfr/ \eqref{adaptedfr}. 

An expression for the Dirichlet Hamiltonian boundary term \eqref{gDH}
in terms of the standard ADM canonical variables associated to $\Sigma$,
and its relation to quasilocal quantities of Brown and York
\cite{Brown-York1,Brown-York2}, 
will be derived in \secref{covfieldeqs}.

\subsection{Determination of allowed boundary conditions}
\label{generalbc}

A necessary and sufficient condition \cite{Wald-Zoupas} 
on variations $\delta\frame{\mu}{a}$
for existence of a Hamiltonian \eqref{defHfr} 
conjugate to $\tfvec{a}$ on $\Sigma$
is given by the antisymmetrized variation of 
the equation \eqref{bteq} for the boundary term. 
This yields 
\EQ\label{frbceq}
\vol{}{ab} \tfvec{c} \w{abc}(\theta,\delta_1\theta,\delta_2\theta)
|_{\partial\Sigma} 
= \metric{a}{b}\der{b} \tb{a}{}(\xi,\theta;\delta_1\theta,\delta_2\theta)
|_{\partial\Sigma} 
\endEQ
with 
\EQ\label{beq}
\tb{a}{}(\xi,\theta;\delta_1\theta,\delta_2\theta)
= \vol{}{ab} \delta_1 \a{}{b}(\xi,\theta;\delta_2\theta)
- \vol{}{ab} \delta_2 \a{}{b}(\xi,\theta;\delta_1\theta) . 
\endEQ
To begin, we simplify the expression \eqref{wfr} 
for $\w{abc}(\theta,\delta_1\theta,\delta_2\theta)$. 
First, using \Eqref{simpleTfr} for $\T{abc}(\theta,\delta\theta)$
and taking into account the orthogonality $\tfvec{a} \s{}{a}=0$,
we have
\EQs
\frac{1}{16} \vol{}{ab} \tfvec{c} \T{abc}(\theta,\delta\theta)
&&
= \tfvec{c} \t{}{c} \h{\mu}{a} 
\delta( \s{}{\nu} \h{}{b\mu} \covgder{a} \frame{\nu}{b} )
\label{Teq}
\endEQs
through the frame decomposition \eqref{frhdecomp}
and the relation 
$\frcon{a}{\mu\nu}(\theta) 
= \frame{c\mu}{} \covgder{a} \frame{\nu}{c} 
= -\frame{c\nu}{} \covgder{a} \frame{\mu}{c}$. 
Now we substitute the identity
$\covgder{a} \frame{\nu}{b}
= \h{a}{c} \covgder{c} \frame{\nu}{b}
- \s{}{a} \s{c}{} \covgder{c} \frame{\nu}{b}$
and then use the relations
$\h{}{a\mu} \delta\s{}{a} =0$, $\h{}{a\mu} \s{}{a} = 0$
to simplify the term 
\EQ
\h{\mu}{a} \delta( \s{}{\nu} \s{}{a} \h{}{b\mu} 
\s{c}{}\covgder{c} \frame{\nu}{b} )
= 0 . 
\endEQ
Thus, \Eqref{Teq} becomes 
\EQ
\frac{1}{16} \vol{}{ab} \tfvec{c} \T{abc}(\theta,\delta\theta)
= \tfvec{c} \t{}{c} \h{}{a\mu}
\delta( \s{}{\nu} \h{\mu}{b} \h{a}{c} \covgder{c} \frame{\nu}{b} ) . 
\endEQ
Finally, we substitute 
$\tfvec{c} \t{}{c}
= \frac{1}{2} \vol{}{ab} \tfvec{c} \vol{abc}{}(h)$. 

Hence, we obtain
\EQ\label{covTfr}
\Pr{\bcsurface} \T{abc}(\theta,\delta\theta)
= 8 \vol{abc}{}(h) \h{\mu}{d} \delta\K{d}{\mu}
\endEQ
where we define
\EQ\label{frK}
\K{a}{\mu}
= \s{}{\nu} \h{}{b\mu} \h{a}{c} \covgder{c} \frame{\nu}{b}
= \s{}{\nu} \h{a}{c} \frcon{c}{\mu\nu}(\theta) . 
\endEQ
Note that
$\s{a}{} \K{a}{\mu} =0$ and $\s{}{\mu} \K{a}{\mu} =0$.
From \Eqref{covTfr}, 
by taking an antisymmetric variation 
and then using \Eqref{varvolhfr} for the variation of $\vol{abc}{}(h)$, 
we have 
\EQs
&&
\Pr{\bcsurface} \w{abc}(\theta,\delta_1\theta,\delta_2\theta)
\nonumber\\&&
=  8\vol{abc}{}(h) \bigg(
( \delta_1\h{\mu}{d} - \h{\mu}{d} \h{e}{\nu} \delta_1\h{\nu}{e} )
\delta_2\K{d}{\mu}
- ( \delta_2\h{\mu}{d} - \h{\mu}{d} \h{e}{\nu} \delta_2\h{\nu}{e} )
\delta_1\K{d}{\mu}
\bigg) . 
\label{wsymplecticdata}
\endEQs
Substitution of this expression into equation \eqref{frbceq} yields
the following result.

\Proclaim{ Theorem 3.4. }{
A Hamiltonian conjugate to $\tfvec{a}$ on $\Sigma$ 
exists for variations $\delta\frame{\mu}{a}$ 
with support on $\partial\Sigma$ 
if and only if
\EQs
&&
\bigg(
( \delta_1\h{\mu}{d}  - \h{\mu}{d} \h{e}{\nu} \delta_1\h{\nu}{e} )
\delta_2\K{d}{\mu}
- ( \delta_2\h{\mu}{d}  - \h{\mu}{d} \h{e}{\nu} \delta_2\h{\nu}{e} )
\delta_1\K{d}{\mu}
\bigg) \bigg|_{\partial\Sigma}
\nonumber\\&&
= \frac{1}{16N} \metric{a}{b}\der{b} 
\tb{a}{}(\xi,\theta;\delta_1\theta,\delta_2\theta)
|_{\partial\Sigma} 
\label{generalbcdeteqfr}
\endEQs
for some 
$\tb{a}{}(\xi,\theta;\delta_1\theta,\delta_2\theta)$
of the form \eqref{beq}. 
The Hamiltonian is given by 
\EQ\label{fullHfr}
H_\Sigma(\xi;\theta)
= 8 \int_\Sigma \tfvec{e} \n{}{\mu}
( \curv{e}{\mu}(\theta) -\frac{1}{2}\frame{\mu}{e} \scurv(\theta) ) d\Sigma
+H_B(\xi;\theta)
\endEQ
with \bdt/ 
\EQ
H_B(\xi,\theta) 
= \int_{\partial\Sigma} \Q{ab}(\xi,\theta) 
-\tfvec{c}\B{abc}(\theta)
\endEQ
where $\B{abc}(\theta)$ is determined from the equation
\EQ\label{Bfr}
\tfvec{c}( \Pr{\bcsurface} \delta\B{abc}(\theta)
- 8 \vol{abc}{}(h) \h{\mu}{d} \delta\K{d}{\mu} )
=  \Pr{\bcsurface} \der{[a} \a{}{b]}(\xi,\theta;\delta\theta) . 
\endEQ
}

Thus, equation \eqref{generalbcdeteqfr}
determines the allowed \bdc/s on variations $\delta\frame{\mu}{a}$
for existence of a Hamiltonian formulation \eqref{fullHfr}
for the vacuum Einstein equations. 
To proceed, we now parallel the analysis of 
the similar \bdc/ determining equation for the Maxwell equations
in \secref{maxwellbc}. 

Two obvious solutions of the determining equation \eqref{generalbcdeteqfr}
with $\tb{a}{}(\xi,\theta;\delta_1\theta,\delta_2\theta)=0$
are given by 
$\delta\h{\mu}{a}|_{\partial\Sigma_t} =0, t\geq 0$, 
which is the Dirichlet \bdc/ \eqref{intrinsicDbcfr}
already considered; 
and by 
\EQ\label{Nbcfr}
\delta\K{a}{\mu} |_{\partial\Sigma_t} =0 ,\quad
t\geq 0
\endEQ
which we call the {\it Neumann \bdc/. } 
For the \bdc/ \eqref{Nbcfr}, 
it follows from \Eqrefs{defHfr}{Bfr}
that the corresponding Hamiltonian \bdt/ is given by 
\EQ\label{HNfr}
\H{N}(\xi;\theta)
= 8\int_{\partial\Sigma} \tfvec{c} \P{N}{c}(\theta) dS
\endEQ
where
\EQ\label{PNfr}
\P{\rm N}{c}(\theta)
= \t{}{\mu} \s{}{\nu} \frcon{c}{\mu\nu}(\theta) 
= \t{a}{} \s{}{\nu} \covgder{c} \frame{\nu}{a}
\endEQ
by a derivation similar to \Eqref{PDfr}. 
In the \onfr/ \eqref{adaptedfr} adapted to 
the boundary 2-surfaces $\partial\Sigma_t$,
we have
\EQ\label{PN}
\P{\rm N}{c}(\vartheta)
= \t{a}{} \covgder{c} \s{}{a}
\endEQ
We refer to this as the {\it Neumann symplectic vector} 
associated to the boundary 2-surfaces $\partial\Sigma_t$. 
Moreover, in this frame the Neumann \bdc/ \eqref{Nbcfr}
becomes
\EQ
\delta\K{a}{\mu} |_{\partial\Sigma_t} 
= \delta( \h{}{b\mu} \K{ab}{} )|_{\partial\Sigma_t} =0 ,\quad
t\geq 0 
\endEQ
in terms of 
\EQ
\K{ab}{}= \h{a}{c} \h{b}{d} \covgder{c} \s{}{d}
\endEQ
which is the extrinsic curvature of
the timelike boundary hypersurface $\bcsurface$ in $(M,g)$. 
Thus, geometrically, the Neumann \bdc/ corresponds to 
fixing the frame components of the boundary hypersurface extrinsic curvature,
\EQ
\delta( \h{}{b\mu} \h{a}{c} \covgder{c} \s{}{b} ) 
=0 
\quad\eqtext{ on $\bcsurface$. }
\endEQ
These components measure the rotation and boost of 
the hypersurface normal $\s{}{a}$ with respect to the frame $\h{a}{\mu}$
under displacement on $\bcsurface$. 

We now investigate more general \bdc/s. 
Note that, 
on the left-side of the determining equation \eqref{generalbcdeteqfr}, 
$\vol{}{ab} \tfvec{c} \w{abc}(\theta,\delta_1\theta,\delta_2\theta)$
involves only the field variations 
$\Pr{\bcsurface} \delta\h{\mu}{a}$ 
and 
$\Pr{\bcsurface}\delta\K{a}{\mu} 
= \Pr{\bcsurface} \delta( \s{}{\nu} \h{a}{c} \frcon{a}{\mu\nu}(\theta) )$.
We call $\h{\mu}{a}$ and $\K{a}{\mu}$ 
the {\it symplectic boundary data}
at $\partial\Sigma_t$
and consider \bdc/s of the form 
\EQ\label{hKbc}
\delta\bcdata{a}{\mu}(\h{c}{\nu},\K{c}{\nu})
|_{\partial\Sigma_t} =0, t\geq 0
\endEQ
where $\bcdata{a}{\mu}(\h{c}{\nu},\K{c}{\nu})$ is locally constructed 
as an algebraic expression in terms of 
the symplectic boundary data and fixed quantities
(including the spacetime coordinates). 
We call \eqref{hKbc} a {\it mixed Dirichlet-Neumann} \bdc/
if $\bcdata{a}{\mu}(\h{c}{\nu},\K{c}{\nu})$ 
is a constant-coefficient linear combination of the parts 
$\Pr{\partial\Sigma}\h{a}{\mu}$, $\Pr{t}\h{a}{\mu}$, 
$\Pr{\partial\Sigma}\K{a}{\mu}$, $\Pr{t}\K{a}{\mu}$
of the Dirichlet and Neumann boundary data in 
\eqrefs{intrinsicDbcfr}{Nbcfr}. 
Here the projections with respect to 
$\Pr{\partial\Sigma}$ and $\Pr{t}$
remove all components proportional to $\s{}{a}$. 

An analysis of the \bdc/ determining equation \eqref{generalbcdeteqfr}, 
given later, leads to the following main results. 

\Proclaim{ Theorem 3.5. }{
The only allowed mixed Dirichlet-Neumann \bdc/s 
for existence of a Hamiltonian \eqref{defHfr}
conjugate to $\tfvec{a}$ on $\Sigma$
are given by 
\EQ\label{mixedDNbcfr}
\Pr{\bcsurface}( a_0 \delta\K{a}{\mu} + b_0 \delta\h{a}{\mu} )
|_{\partial\Sigma_t} =0, \quad 
t\geq 0, 
\endEQ
or equivalently
\EQ\label{mixedbdfr}
\bcdata{a}{\mu}(\h{c}{\nu},\K{c}{\nu})
= a_0 \K{a}{\mu} + b_0 \h{a}{\mu}
\endEQ
for constants $a_0,b_0$ 
(and $\b{a}{}=0$ in \Eqref{generalbcdeteqfr}).
In the cases $a_0=0$ or $b_0=0$, 
respectively Dirichlet or Neumann \bdc/s,
the corresponding Hamiltonian \bdt/s are given by 
\Eqrefs{HDfr}{PDfr}, and \Eqrefs{HNfr}{PNfr}. 
In the case $a_0 \neq 0, b_0 \neq 0$,
the corresponding Hamiltonian \bdt/ is given by 
\EQ\label{HDNfr}
\H{DN}(\xi;\theta)
= 8\int_{\partial\Sigma} \tfvec{c} \P{}{c}(\theta) dS
\endEQ
where, now, 
\EQ\label{PNbcfr}
\P{}{c}(\theta)
= \P{N}{c}(\theta)+6 \frac{b_0}{a_0} \t{}{c} . 
\endEQ
(Note, the \bdt/s here are unique up to adding
an arbitrary covector function of the boundary data \eqref{mixedbdfr}
to $\P{}{c}(\theta)$.)
}

A similar covariant derivation of 
the pure Dirichlet and Neumann boundary terms 
is presented in \Ref{Nester1,Nester2} from a different perspective. 
In Theorem~3.5, note that 
\Eqref{mixedDNbcfr} represents a one-parameter $a_0/b_0$ family of \bdc/s. 
In particular, 
in contrast to the two-parameter family of analogous
mixed Dirichlet-Neumann \bdc/s \eqref{MEbc}
allowed for the Maxwell equations,
here decompositions of the symplectic boundary data
with respect to $\Pr{\partial\Sigma}$ and $\Pr{\xi}$
do not yield \bdc/s satisfying 
the determining equation \eqref{generalbcdeteqfr}.

The form of the mixed Dirichlet-Neumann \bdc/ \eqref{mixedDNbcfr}
suggests we also consider \bdc/s specified by 
a trace part and trace-free part with respect to 
the boundary hypersurface frame $\h{a}{\mu}$:
\EQ\label{othhKbc}
\delta\othbcdata{}{}(\h{c}{\nu},\K{c}{\nu})
|_{\partial\Sigma_t} =0 ,\quad
\delta\othbcdata{a}{\mu}(\h{c}{\nu},\K{c}{\nu})
|_{\partial\Sigma_t} =0 ,\quad
t\geq 0
\endEQ
with $\h{\mu}{a}\othbcdata{a}{\mu}(\h{c}{\nu},\K{c}{\nu})=0$. 
Taking the trace of the symplectic boundary data variations yields
\EQ
\h{\mu}{a} \delta\K{a}{\mu} 
= \delta\trK +\h{\mu}{b} \h{\nu}{a} \K{a}{\mu} \delta\h{b}{\nu}
\endEQ
and 
\EQ
\h{\mu}{a} \delta\h{a}{\mu}
= \delta \ln|h|
\endEQ
where 
$\trK=\h{\mu}{a}\K{a}{\mu}$
is the trace of $\K{a}{\mu}$
and 
$h=\det(\h{a}{\mu})$
is the determinant of the frame components $\h{a}{\mu}$
in local coordinates. 

\Proclaim{ Theorem 3.6. }{
Allowed \bdc/s \eqref{othhKbc}
for the existence of a Hamiltonian \eqref{defHfr}
conjugate to $\tfvec{a}$ on $\Sigma$
are given by 
\EQs
&&
\Pr{\bcsurface} \delta( \K{a}{\mu} -\frac{1}{3} \h{a}{\mu} \trK )
|_{\partial\Sigma_t} =0 ,\quad
t\geq 0,
\label{othbcfr}\\
&&
( a_0 \delta\trK + b_0 \delta\ln|h| )
|_{\partial\Sigma_t} =0 ,\quad
t\geq 0,
\label{othbc'fr}
\endEQs
or equivalently
\EQ\label{othmixedbdfr}
\othbcdata{}{}(h,K) 
= a_0 \trK + b_0 \ln|h| ,\quad
\othbcdata{a}{\mu}(\h{c}{\nu},\K{c}{\nu}) 
= \Pr{\bcsurface}( \K{a}{\mu} -\frac{1}{3} \h{a}{\mu} \trK )
\endEQ
for constants $a_0,b_0$ 
(and $\b{a}{}=0$ in \Eqref{generalbcdeteqfr}).
The corresponding Hamiltonian \bdt/ is given by 
\EQ\label{Hothfr}
H_B(\xi,\theta)
= 8\int_{\partial\Sigma} \tfvec{c} 
( \frac{1}{3}\P{D}{c}(\theta) + \frac{2}{3}\P{N}{c}(\theta) 
+4\frac{b_0}{a_0} \t{}{c} ) dS
\endEQ
(which is unique up to adding a term 
depending on an arbitrary covector function of the boundary data 
\eqref{othmixedbdfr}).
}

Finally, we remark that the mixed \bdc/s in Theorems~3.5 and~3.6
admit the following two generalizations. 

First, 
\EQs
&&
\othbcdata{a}{\mu}(\h{c}{\nu},\K{c}{\nu}) 
= a(x,\trK,\ln|h|) 
\Pr{\bcsurface}( \K{a}{\mu} -\frac{1}{3} \h{a}{\mu} \trK ),
\label{generalmixedbc}\\
&&
\othbcdata{}{}(h,K) 
= b(x,\trK,\ln|h|) ,\quad
\der{\trK} b \neq 0
\endEQs
for arbitrary functions $a(x,\trK,\ln|h|),b(x,\trK,\ln|h|)$. 

Second, 
\EQ\label{generalmixedbc'}
\bcdata{a}{\mu}(\h{c}{\nu},\K{c}{\nu})
= a(x,\trK,\ln|h|)
\Pr{\bcsurface}( \K{a}{\mu} +b(x,\trK,\ln|h|) \h{a}{\mu}  ) ,\quad
b\neq -\frac{1}{3} \trK
\endEQ
for an arbitrary function $b(x,\trK,\ln|h|)$, 
with the function $a(x,\trK,\ln|h|)$ 
now satisfying the linear partial differential equation
\EQ
(\der{\trK} \hat b) \der{\ln|h|} a
+ (\hat b + \der{\ln|h|} \hat b) \der{\trK} a
= \frac{2}{9} (-1+3\der{\trK} \hat b) a, \quad
\hat b = b +\frac{1}{3} \trK
\endEQ
obtained from the determining equation \eqref{generalbcdeteqfr}. 
The general form of $a$ is given by solving 
the characteristic ordinary differential equations
\EQ
\frac{d\ln|h|}{\der{\trK} \hat b}
= \frac{d\trK}{\hat b + \der{\ln|h|} \hat b}
= \frac{d a}{\frac{2}{9} (-1+3\der{\trK} \hat b) a}
\endEQ
in terms of the variables $\ln|h|,\trK,a$. 
For instance, if $b$ is taken to be linear homogeneous in $\trK$,
then one has
$b=\lambda \trK$, 
$a=f(x,\lambda \trK\ln|h|) |h|^{\frac{2}{3}\frac{\lambda}{3\lambda -1}}$,
where $\lambda=\const$
and $f$ is an arbitrary function. 

\Proclaim{ Proofs of Theorems: }{}
Since any \bdc/ locally constructed from the symplectic boundary data
is linear homogeneous in 
$\Pr{\bcsurface} \delta\h{a}{\mu}$ 
and $\Pr{\bcsurface} \delta\K{a}{\mu}$,
we begin by finding all such solutions of 
the determining equation \eqref{generalbcdeteqfr}. 

First we show that $\tb{a}{}=0$. 
The right-side of \Eqref{generalbcdeteqfr} necessarily involves
terms with at least one derivative on $\delta\frame{\mu}{a}$,
while only first-order derivatives of $\delta\frame{\mu}{a}$ 
appear on the left-side of \Eqref{generalbcdeteqfr}
through 
\EQ\label{varKfr}
\Pr{\bcsurface}\delta\K{a}{\mu}
= \s{}{\nu} \h{a}{c} \delta\frcon{c}{\mu\nu}(\theta)
\endEQ
due to \Eqrefs{frK}{varhid}. 
Thus, for a balance in numbers of derivatives,
we must have
\EQ\label{aeq}
\a{}{a} = \a{\ b}{a\mu}(\theta) \delta\frame{\mu}{b} 
\endEQ
for some $\a{\ b}{a\mu}(\theta)$ locally constructed out of
$\frame{\mu}{a}$ and fixed quantities. 
This yields, for the antisymmetrized variation of $\a{}{a}$,
\EQ\label{baeq}
\b{}{a} 
= \a{\ bc}{a\mu\nu}(\theta) \delta_1\frame{\mu}{b} \delta_2\frame{\nu}{c}
\endEQ
where
$\a{\ bc}{a\mu\nu}(\theta) = -\a{\ cb}{a\nu\mu}(\theta)$ is the curl of 
$\a{\ b}{a\mu}(\theta)$ with respect to $\frame{\nu}{c}$. 
Then, using \Eqrefs{beq}{frconeq}, 
we collect all the terms on the left-side of \Eqref{generalbcdeteqfr}
linearly independent of $\delta\h{a}{\mu}$, $\delta\K{a}{\mu}$. 
Through \Eqrefs{varKfr}{frcon}, 
the coefficients of these terms yield
\EQ\label{asymmskeweq}
\s{}{c} \ta{(be)c}{\ \;\mu\nu}(\theta) =0 ,\quad
\s{}{b} \ta{[be]c}{\ \;\mu\nu}(\theta) =0 ,\quad
\s{}{c} \ta{[be]c}{\ \;\mu\nu}(\theta) =0 
\endEQ
where
\EQ\label{taeq}
\ta{ebc}{\ \mu\nu}(\theta)
= \vol{}{ae} \a{\ bc}{a\mu\nu}(\theta) 
= -\ta{ecb}{\ \nu\mu}(\theta) . 
\endEQ
These algebraic equations are straightforward to solve, 
leading to 
\EQ\label{aform}
\a{abc}{\ \mu\nu}(\theta)
= \t{[b}{} \asub{1}{c]a}{\mu\nu}(\theta)
+ \asub{2}{abc}{\mu\nu}(\theta)
+ \t{b}{} \t{c}{} \asub{3}{a}{\mu\nu}(\theta)
+ \t{(b}{} \asub{4}{c)a}{\mu\nu}(\theta)
\endEQ
for some 
\EQ
\asub{1}{ca}{\mu\nu} 
= \metric{b}{c} \metric{d}{a} \asub{1}{bd}{(\mu\nu)} ,\
\asub{2}{abc}{\mu\nu}
=-\metric{e}{a} \metric{d}{b} \metric{f}{c} \asub{2}{edf}{\nu\mu} ,\
\asub{3}{a}{\mu\nu}
= \metric{e}{a} \asub{3}{e}{[\mu\nu]} ,\
\asub{4}{ca}{\mu\nu}
= \metric{b}{c} \metric{d}{a} \asub{4}{bd}{[\mu\nu]} . 
\label{acoeff}
\endEQ
Then, returning to \Eqref{aeq}, 
we note that
$\a{\ b}{a\mu}(\theta) 
= \frame{\alpha}{a} \frame{b}{\beta} \a{\beta}{\mu\alpha}$
for some $\a{\beta}{\mu\alpha}$ 
that is locally constructed only from fixed quantities
since it is a scalar expression. 
Thus we immediately obtain
\EQ\label{aform'}
\a{abc}{\ \mu\nu}(\theta)
= \frame{b}{\beta} \frame{c}{\gamma} \frame{a\alpha}{}
( \a{\beta}{\nu\mu} \id{\alpha}{\gamma} 
- \a{\gamma}{\mu\nu} \id{\alpha}{\beta} 
- \a{\gamma}{\alpha\mu} \id{\nu}{\beta} 
+ \a{\beta}{\alpha\mu} \id{\nu}{\gamma} ) . 
\endEQ
By equating expressions \eqrefs{aform}{aform'},
we find that after some algebraic analysis
\EQ
\a{\gamma}{\nu\mu} = \id{\nu}{\gamma} \asub{0}{}{\mu} 
\label{solvea}
\endEQ
for some $\asub{0}{}{\mu}$. 
Now, substitution of expression \eqref{solvea} 
back into \Eqrefs{aform}{aform'}
easily leads to the result
\EQ
\asub{1}{ca}{\mu\nu} 
= \asub{2}{abc}{\mu\nu} 
= \asub{3}{a}{\mu\nu} 
= \asub{4}{ca}{\mu\nu}
=0 . 
\endEQ
Hence, from \Eqref{aform}, we have
$\a{\ bc}{a\mu\nu}(\theta)=0$
and so $\b{}{a}=0$. 
This establishes that $\tb{a}{}=0$. 

Consequently, 
the determining equation \eqref{generalbcdeteqfr}
reduces to 
\EQ
( \delta_1\h{\mu}{d} - \h{\mu}{d} \h{e}{\nu} \delta_1\h{\nu}{e} )
\delta_2\K{d}{\mu}
- ( \delta_2\h{\mu}{d} - \h{\mu}{d} \h{e}{\nu} \delta_2\h{\nu}{e} )
\delta_1\K{d}{\mu}
=0, 
\endEQ
which is equivalent to 
\EQ
\h{[\nu}{d} \h{\mu]}{e} 
( \delta_1\h{e}{\nu} \delta_2\K{d}{\mu}
- \delta_2\h{e}{\nu} \delta_1\K{d}{\mu} )
=0 . 
\label{simplebcdeteqfr}
\endEQ
Then the algebraic solution of \Eqref{simplebcdeteqfr}
in terms of $\Pr{\bcsurface} \delta\h{e}{\nu}$ 
and $\Pr{\bcsurface} \delta\K{d}{\mu}$
has the form
\EQ\label{generalbcform}
\coeff{1}{a \nu}{\mu b} \Pr{\bcsurface}\delta\h{b}{\nu}
+ \coeff{2}{a \nu}{\mu b} \Pr{\bcsurface}\delta\K{b}{\nu}
=0
\endEQ
for some coefficient tensors $\coeff{}{a \nu}{\mu b}$
such that 
\EQ
\coeff{}{a \nu}{\mu b}
\h{[\alpha}{a} \h{\mu]}{c} 
= \coeff{}{a \alpha}{\mu c}
\h{[\nu}{a} \h{\mu]}{b} . 
\label{coeffeq}
\endEQ
It is straightforward to show that \Eqref{coeffeq} 
holds iff
\EQ
\coeff{}{a \nu}{\mu b} = 
\othcoeff{}{a \nu}{\mu b} 
- \h{a}{\mu} \othcoeff{}{\nu}{b} + \h{\nu}{b} \othcoeff{}{a}{\mu} 
\label{coeffeq'}
\endEQ
where
$\othcoeff{}{a \nu}{\mu b} \h{\alpha}{a} \h{\mu}{c} 
= \othcoeff{}{a \alpha}{\mu c} \h{\nu}{a} \h{\mu}{b}$
is the symmetric part of $\coeff{}{a \nu}{\mu b}$
in the index pairs $(a,\mu)$ and $(b,\nu)$,
and where
$\othcoeff{}{\nu}{b} 
= \h{\mu}{a} \othcoeff{}{a \nu}{\mu b},
\othcoeff{}{a}{\mu} 
= \h{b}{\nu} \othcoeff{}{a \nu}{\mu b}$
is the trace part of $\coeff{}{a \nu}{\mu b}$
in the frame $\h{a}{\mu}$. 
Thus, we have established the following result.

\Proclaim{ Lemma 3.7. }{
All solutions of the determining equation \eqref{generalbcdeteqfr}
for allowed \bdc/s that are linear homogeneous in 
$\Pr{\bcsurface} \delta\h{a}{\mu}$ 
and $\Pr{\bcsurface} \delta\K{a}{\mu}$
have the form \eqref{generalbcform}
where the coefficient tensors are given by \Eqref{coeffeq'}.
}

Now, for mixed Dirichlet-Neumann \bdc/s, we take
\EQs
&&
\coeff{1}{a \nu}{\mu b} 
= a_1 \Prop{\partial\Sigma}{a}{b} \metric{\nu}{\mu}
+ a_0 \Prop{t}{a}{b} \t{}{\nu} \t{\mu}{}
\label{coeffsol}\\&&
\coeff{2}{a \nu}{\mu b} 
=  b_1 \Prop{\partial\Sigma}{a}{b} \metric{\nu}{\mu}
+ b_0 \Prop{t}{a}{b} \t{}{\nu} \t{\mu}{}
\label{coeffsol'}
\endEQs
where $a_0,a_1,b_0,b_1$ are constants. 
Then the requirement \eqref{coeffeq'}
leads directly to 
\EQ\label{coeffs}
a_0=a_1 ,\ b_0=b_1 . 
\endEQ
Substitution of \Eqsref{coeffsol}{coeffs}
into \Eqref{generalbcform} 
yields the mixed Dirichlet-Neumann \bdc/s
\eqref{mixedDNbcfr}. 

Finally, the other \bdc/s \eqrefs{othbcfr}{othbc'fr}
arise from 
\EQs
&&
\coeff{1}{a \nu}{\mu b} 
= \frac{1}{2} \frac{b_0}{a_0} \h{a}{\mu} \h{\nu}{b} 
-\frac{1}{3} \trK \h{d}{e} \h{\nu}{\mu} , 
\\&&
\coeff{2}{a \nu}{\mu b} 
= \h{a}{b} \h{\nu}{\mu} , 
\endEQs
which are easily verified to satisfy the requirement \eqref{coeffeq'}.

This completes the proofs of Theorems~3.5 and~3.6. 
\endproof

As a concluding remark, we note that Lemma~3.7 yields 
the following necessary and sufficient determining equations 
for finding all \bdc/s \eqref{hKbc}.

\Proclaim{ Lemma 3.8. }{
All allowed \bdc/s of the form 
$\bcdata{a}{\mu}(\h{c}{\nu},\K{c}{\nu})|_{\partial\Sigma} =0$ 
for existence of a Hamiltonian conjugate to $\tfvec{a}$ on $\Sigma$
are given by the solutions of the equations
\EQs
&&
\Parder{\bcdata{a}{\mu}}{\h{b}{\nu}} 
\h{[\alpha}{a} \h{\mu]}{c} 
= \Parder{\bcdata{a}{\mu}}{\h{c}{\alpha}} 
\h{[\nu}{a} \h{\mu]}{b} , 
\label{PDEbcdeteqfr}
\\&&
\Parder{\bcdata{a}{\mu}}{\K{b}{\nu}} 
\h{[\alpha}{a} \h{\mu]}{c} 
= \Parder{\bcdata{a}{\mu}}{\K{c}{\alpha}} 
\h{[\nu}{a} \h{\mu]}{b} . 
\label{PDEbcdeteqfr'}
\endEQs }

We will leave a general analysis of 
the \bdc/ determining equations \eqrefs{PDEbcdeteqfr}{PDEbcdeteqfr'}
for elsewhere.

\subsection{ Relation between covariant and canonical Hamiltonians 
and boundary terms }
\label{covfieldeqs}

To conclude this section, 
we first give a brief discussion of the Hamiltonian field equations
for General Relativity using the covariant symplectic structure
and Noether charge Hamiltonian in \secref{Dbcanalysis}. 
Then we discuss the boundary terms in the Dirichlet and Neumann Hamiltonians
expressed in standard ADM canonical variables. 

For a Hamiltonian \eqref{defHfr} 
conjugate to $\tfvec{a}$ on $\Sigma$, 
the associated field equations are obtained through 
the presymplectic form \eqref{presymplecticfreq} by 
the variational principle
\EQ\label{GRvarder}
\Omega_\Sigma(\theta,\delta\theta,\Lie{\xi}\theta)
- H'_\Sigma(\xi;\theta,\delta\theta) 
=\int_\Sigma 8\tfvec{d} \n{}{d} ( \curv{\mu}{e}(\theta) 
-\frac{1}{2}\frame{e}{\mu} ) \delta\frame{\mu}{e} \ d\Sigma
=0 
\endEQ
for arbitrary variations $\delta\frame{\mu}{e}|_\Sigma$.
These field equations split into 
evolution equations and constraint equations
with respect to $\Sigma$
corresponding to a decomposition of $\frame{\mu}{a}$ into
dynamical and non-dynamical components 
determined by \cite{Wald-Lee} the degeneracy of 
\EQ\label{symplecticdensity}
\Omega_\Sigma(\theta,\delta_1\theta,\delta_2\theta)
= \int_\Sigma \w{abc}(\theta,\delta_1\theta,\delta_2\theta)
= \int_\Sigma \frac{1}{6} \vol{}{abcd} \n{}{d} 
\w{abc}(\theta,\delta_1\theta,\delta_2\theta) d\Sigma . 
\endEQ
For this purpose, it is convenient to partially fix 
the SO(3,1) local gauge freedom in $\frame{\mu}{e}$ 
analogously to conditions \eqsref{gaugefr}{moregaugefr}
by choosing the frame components
$\n{\mu}{} =\n{a}{} \frame{\mu}{a}$
to be fixed constants on $M$. 
Then, through a simplification of the symplectic current here 
similar to \Eqref{wsymplecticdata}, 
we obtain
\EQ
\frac{1}{48} \vol{}{abcd} \n{}{d} 
\w{abc}(\theta,\delta_1\theta,\delta_2\theta)
= -( \delta_1\q{\mu}{d} - \q{\mu}{d} \q{e}{\nu} \delta_1\q{\nu}{e} )
\delta_2\p{d}{\mu}
+ ( \delta_2\q{\mu}{d} - \q{\mu}{d} \q{e}{\nu} \delta_2\q{\nu}{e} )
\delta_1\p{d}{\mu}
\endEQ
where we define
\EQ
\q{a}{\mu}=\frame{\mu}{a} +\n{}{a}\n{\mu}{} ,\quad
\p{a}{\mu} = \n{}{\nu} \q{}{b\mu} \q{a}{c} \covgder{c} \frame{\nu}{b}
= \n{}{\nu} \q{a}{c} \frcon{c}{\mu\nu}(\theta)
\endEQ
with $\q{ab}{} =\g{ab}{} +\n{}{a}\n{}{b}$,
which are counterparts of $\h{a}{\mu},\K{a}{\mu}$ 
associated to the spacelike hypersurface $\Sigma$. 
(Geometrically, $\q{a}{\mu}$ is a frame for the hypersurface metric 
$\q{ab}{} =\q{a}{\mu}\q{b}{\nu}\flat{\mu\nu}{}$,
while $\p{a}{\mu}$ represents the frame components of 
the hypersurface extrinsic curvature 
$\surfexcurv{ab}{} = \q{a}{\mu}\p{b}{\nu}\flat{\mu\nu}{}
= \q{a}{c}\covgder{c}\n{}{b}$.)
Hence, the degeneracy directions of 
the presymplectic form \eqref{symplecticdensity}
are given by variations $\delta\frame{\mu}{a}$ satisfying
\EQ\label{qpdegeneracy}
\delta(\q{a}{\mu}/|q|) =\delta \p{a}{\mu}=0 . 
\endEQ
where $|q|=det(\q{a}{\mu})$ and $\delta\ln|q| = \q{\nu}{b}\delta\q{b}{\nu}$.
This immediately leads to the following result, 
analogous to the discussion in \secref{MEpreliminaries}
for the Maxwell equations. 

\Proclaim{ Proposition 3.9. }{
For a Hamiltonian $H_\Sigma(\xi;\theta)$
conjugate to $\tfvec{a}$ on $\Sigma$, 
let $\theta_{\ker\omega}$ denote the field components of 
$\frame{\mu}{a}$ and $\frcon{a}{\mu\nu}(\theta)$
invariant under the symplectic degeneracy directions \eqref{qpdegeneracy}, 
and let $\theta_{\omega}$ denote the remaining field components 
modulo $\theta_{\ker\omega}$. 
Then there is corresponding decomposition of the Hamiltonian field equations 
given by 
$H'_\Sigma(\xi;\theta,\delta\theta_{\ker\omega}) =0$
and 
$H'_\Sigma(\xi;\theta,\delta\theta_{\omega}) 
=\Omega_\Sigma(\theta,\delta\theta_{\omega},\Lie{\xi}\theta)$, 
which respectively yield
\EQs
&& 
\n{a}{} \curv{a}{\mu}(\theta) -\frac{1}{2} \n{\mu}{} \scurv(\theta) =0 , 
\label{GRnondyneq}\\
&&
\q{a}{b} \curv{b}{\mu}(\theta) =0 . 
\label{GRdyneq}
\endEQs
These field equations arise equivalently by variation of
$\Pr{n}\frame{\mu}{a}= -\n{}{a} \n{\mu}{}$ 
and $\Pr{\Sigma}\frame{\mu}{a} = \q{a}{\mu}$ 
in the Lagrangian \eqref{Lfr}. 
}

Equations \eqrefs{GRnondyneq}{GRdyneq} are the frame components of
the standard 3+1 split of the vacuum Einstein equations \cite{Wald-book}
into constraint equations and time-evolution equations
for the hypersurface metric $\q{ab}{}$. 
Thus, the Hamiltonian field equations given by 
the variational principle \eqref{GRvarder}
constitute a covariant formulation of 
the standard ADM Hamiltonian equations for General Relativity. 

With respect to the spacelike hypersurface $\Sigma$, 
one has a decomposition of $\tfvec{a}$ into normal and tangential parts
\EQ
\tfvec{a} = \tN{}{} \n{a}{} +\tN{a}{}
\endEQ
where $\tN{a}{}=\Pr{\Sigma}( \tfvec{a} )$
and $\tN{}{}=-\tfvec{a} \n{}{a}$
define the lapse and shift of the time flow vector field $\tfvec{a}$.
By use of the Gauss-Codacci equations, 
we straightforwardly see that 
the volume part of a Hamiltonian \eqref{defHfr} 
conjugate to $\tfvec{a}$ on $\Sigma$ is given by 
the ``pure constraint form'' \cite{Wald-Iyer2}
\EQ
H(\xi;\theta) 
= 4\int_\Sigma 
\tN{}{}( \surfcurv{}{} + \surfexcurv{}{}{}^2 
-\surfexcurv{ab}{} \surfexcurv{}{ab} ) 
+ 2\tN{c}{}( \surfder{}{b} \surfexcurv{bc}{} -\surfder{c}{} \surfexcurv{}{} )
\ d\Sigma 
\endEQ
where $\surfcurv{ab}{}$ and $\surfexcurv{ab}{}$ 
are the Ricci curvature and extrinsic curvature 
of the metric $\Pr{\Sigma}\g{ab}{}=\q{ab}{}$,
$\surfcurv{}{}=\surfcurv{a}{a}$ 
and $\surfexcurv{}{}=\surfexcurv{a}{a}$ 
are the corresponding scalar curvatures,
and $\surfder{a}{}$ is the derivative operator associated with $\q{ab}{}$.
(An analogous result holds more generally for any diffeomorphism covariant
Lagrangian field theory \cite{Wald-Lee}. )
This demonstrates, explicitly, that our covariant analysis of
allowed boundary conditions and corresponding boundary terms 
for General Relativity in \secrefs{Dirichletbc}{generalbc} 
is equivalent to a canonical analysis of the ADM Hamiltonian. 

Now, consider Dirichlet or Neumann boundary conditions imposed 
at the 2-surfaces $\partial\Sigma_t$, for $t\geq 0$. 
On solutions of the Hamiltonian field equations, 
the total Hamiltonian $H_\Sigma(\xi;\theta)$ reduces, respectively, 
to the Dirichlet and Neumann boundary terms \eqrefs{HDfr}{HNfr}. 
Let $\u{a}{}$ denote the outward unit normal to 
$\partial\Sigma_t$ in $\Sigma_t$. 
Let $\adothfr{\mu}{a}$ be an \onfr/ adapted to $\Sigma_t$
given by 
\EQ\label{adaptedothfr}
\adothfr{0}{a} =\n{}{a}, 
\adothfr{1}{a} =\u{}{a},
\adothfr{2}{[a} \adothfr{3}{b]} =\vol{ab}{},
\adothfr{2}{a} =\vol{a}{b} \adothfr{3}{b} 
\endEQ
which is related to the frame $\adfr{\mu}{a}$ adapted to $\bcsurface$ 
by a boost in the normal space $T^\perp(\partial\Sigma_t)$
to the boundary 2-surface $\partial\Sigma_t$,
\EQ\label{boost}
\t{a}{} = \n{a}{} \cosh\chi +\u{a}{} \sinh\chi ,\quad
\s{a}{} = \u{a}{} \cosh\chi +\n{a}{} \sinh\chi .
\endEQ
Through the corresponding boost relation \eqref{changefrcon}
applied to the symplectic vectors \eqrefs{PD}{PN}, 
the Hamiltonian boundary terms take the respective form
\EQs
&& 
\tfvec{c} \P{\rm D}{c}(\tilde\vartheta)
= \tfvec{c} ( \n{a}{} \metric{c}{d} \covgder{d} \u{}{a} 
- \n{}{c} \metric{}{bd}\covgder{b} \u{}{d}
+ \u{}{c} \metric{}{bd}\covgder{b} \n{}{d}
-  \metric{c}{d}\covgder{d} \chi ) , 
\label{PDboost}\\
&&
\tfvec{c} \P{\rm N}{c}(\tilde\vartheta)
= \tfvec{c} ( \n{a}{} \covgder{c} \u{}{a} - \covgder{c} \chi ) .
\label{PNboost}
\endEQs
These expressions can be simplified in terms of 
the hypersurface metric $\q{ab}{}$, 
extrinsic curvature $\surfexcurv{ab}{}$,
and acceleration $\surfaccel{b}=\n{e}{} \covgder{e} \n{}{b}$. 
We find that
\EQs
&& 
\tfvec{c} \P{\rm D}{c}(\tilde\vartheta)
= \tN{}{} \trexcurv -\tN{a}{} \u{b}{} \surfexcurv{ab}{}
- \tN{a}{\parallel} \der{a} \chi ,
\label{PDADM}\\
&&
\tfvec{c} \P{\rm N}{c}(\tilde\vartheta)
= -\tN{}{} \u{b}{}\surfaccel{b} -\tN{a}{} \u{b}{} \surfexcurv{ab}{}
- \tN{}{} \der{t} \chi - \tN{}{\perp} \u{a}{}\der{a} \chi 
-  \tN{a}{\parallel} \der{a} \chi ,
\label{PNADM}
\endEQs
where $\tN{a}{\parallel} =\Pr{\partial\Sigma}( \tN{a}{} )$, 
$\tN{}{\perp} = \u{}{a} \tN{a}{}$
are the tangential and normal parts of the shift 
with respect to $\partial\Sigma_t$, 
and $\trexcurv = \metric{}{ab} \covgder{a} \u{}{b}$ 
is the mean extrinsic curvature of $\partial\Sigma_t$ in $\Sigma_t$. 

We note that this form of the Dirichlet and Neumann boundary terms 
\eqrefs{PDADM}{PNADM} 
agrees with the canonical analysis of boundary terms 
for the ADM Hamiltonian carried out in \Refs{Brown-York1,Brown-York2}. 
Moreover, in the case when $\Sigma_t$ is orthogonal to $\bcsurface$, 
\ie/ $\chi=0$, 
the surface integral 
$\int_{\partial\Sigma} \tfvec{c} \P{\rm D}{c}(\tilde\vartheta) dS$
for suitable choice of $\tfvec{c}$ reproduces Brown and York's 
expressions for quasilocal 
energy, normal momentum, and tangential momentum quantities
(respectively, 
$\tN{}{}=1,\tN{a}{}=0$; 
$\tN{}{}=0,\tN{a}{\parallel}=0,\tN{}{\perp}=1$; 
$\tN{}{}=\tN{}{\perp}=0,\tN{a}{\parallel}\neq 0$). 
Further discussion of quasilocal quantities associated to 
the Dirichlet and Neumann symplectic vectors \eqrefs{PD}{PN}
will be left for elsewhere.

\section{ Concluding remarks }
\label{conclusion}

In this paper we have given a mathematical investigation of
\bdc/s on the gravitational field 
required for the existence of a well-defined covariant Hamiltonian 
variational principle for General Relativity
when spatial boundaries are considered,
with a fixed time-flow vector field. 
In particular, 
a main result is that 
we obtain a covariant derivation of 
Dirichlet, Neumann, and mixed type \bdc/s for the gravitational field 
in any fixed spatially bounded region of spacetime. 
We show that the resulting Dirichlet and Neumann Hamiltonians
lead to covariant Hamiltonian field equations
which are equivalent to the standard 3+1 split of the Einstein equations
into constraint equations and time-evolution equations. 
In addition, we obtain a uniqueness result for the allowed \bdc/s 
based on the covariant symplectic structure 
associated to the Einstein equations. 

However, we do not address the purely analytical issue of 
whether the boundary-initial value problem for the Einstein equations
is well-posed with these \bdc/s
(\ie/ do there exist solutions of the Einstein equations
satisfying the \bdc/s, initial conditions, and constraints). 
For work in that direction, see \eg/ \Ref{Friedrich}. 

A further interesting generalization of our results would be to treat 
a spacetime region whose spatial boundary is dynamical
\eg/ a black-hole horizon or Cauchy boundary. 
We note that \bdc/s for this situation may be investigated 
by allowing the time-flow vector field to depend on the spacetime metric
instead of being a fixed quantity. 
This analysis will be pursued elsewhere.

\acknowledgments
The authors thank the organizers of the 2nd workshop on Formal Geometry
and Mathematical Physics where this work was initiated. 
Bob Wald is thanked for helpful discussions.

\appendix
\section*{Noether Charge Method}
\label{formalism}

First of all, 
consider in $n$ spacetime dimensions
a general Lagrangian field theory 
for a set of fields denoted collectively by $\phi$. 
It will be assumed that these fields are defined as
sections of a vector bundle $E$ 
over the spacetime manifold $M$,
using local coordinates on $M$ and $E$. 
The theory will be assumed to be described by 
a Lagrangian $n$-form $L(\phi)$
that is locally constructed 
out of the fields $\phi$ and their partial derivatives 
$\partial^k\phi$ to some finite order $k$
(and fixed background structure, if any, on $M$ and $E$). 

The Lagrangian $L(\phi)$ provides a variational principle
\EQ\label{action}
S(\phi) = \int_M L(\phi)
\endEQ
which yields the field equations $\E(\phi)=0$
obtained as the stationary points of $S(\phi)$, 
\EQ
\delta S(\phi) = \int_M \delta L(\phi) 
=  \int_M \E(\phi)\delta\phi =0
\endEQ
under variations $\delta\phi$ of $\phi$
with compact support on $M$.
For arbitrary variations $\delta\phi$, 
which are not restricted to have compact support, 
one then has a variational identity
\EQ\label{varL}
L'(\phi,\delta\phi)\equiv
\delta L(\phi)= \E(\phi) \delta\phi  + d \Theta(\phi,\delta\phi)
\endEQ
where $\Theta (\phi,\delta\phi)$ is an $(n-1)$-form, 
called the {\it symplectic potential},
derived through formal integration by parts. 
This yields a well-defined locally constructed formula 
for $\E(\phi)$ and $\Theta(\phi,\delta\phi)$ 
in terms of $\phi$, $\delta\phi$,
and their partial derivatives to a finite order.
The symplectic potential is used to define 
the {\it presymplectic form}
on a fixed 
hypersurface $\Sigma$, 
\EQ
\Omega_\Sigma(\phi, \delta_1\phi, \delta_2\phi)
\equiv \int_\Sigma\omega(\phi,\delta_1\phi, \delta_2\phi)
\endEQ
in terms of the {\it symplectic current} $(n-1)$-form $\omega$ 
given by
\EQ
\omega(\phi,\delta_1\phi,\delta_2\phi) \equiv
\delta_1\Theta(\phi,\delta_2\phi)- \delta_2\Theta(\phi,\delta_1\phi) . 
\endEQ
The symplectic current satisfies 
$d\omega(\phi,\delta_1\phi,\delta_2\phi)= 
\E'(\phi,\delta_2\phi)\delta_1\phi
-\E'(\phi,\delta_1\phi)\delta_2\phi$
with $\E'(\phi,\delta\phi) \equiv \delta \E(\phi)$. 
Hence, $\omega(\phi,\delta_1\phi,\delta_2\phi)$ is closed
for variations on solutions, 
\EQ
d\omega(\Phi,\delta_1\Phi,\delta_2\Phi)= 0
\endEQ
where $\Phi$ denotes $\phi$ restricted to satisfy $\E(\phi)=0$,
and $\delta\Phi$ denotes $\delta\phi$ restricted to satisfy 
$\E'(\Phi,\delta\phi)=0$, 
\ie/ $\delta\Phi$ is, formally, 
a tangent vector field on the space of solutions. 
Consequently, 
$\Omega_\Sigma(\phi, \delta_1\phi, \delta_2\phi)$
is unchanged by deformations of 
the spacelike surface $\Sigma$ 
in any compact region of $M$. 

In the previous constructions, 
a change of coordinates on $M$ or $E$ leaves 
$\E(\phi)$ unchanged \cite{Anderson}, 
while $\Theta(\phi,\delta\phi)$ changes in general 
by an exact locally constructed $(n-1)$-form $d\nu(\phi,\delta\phi)$.
However, 
one can show that \cite{Wald-Lee}
if the Lagrangian $L(\phi)$ is at most second order 
in partial derivatives $\partial^k\phi$ ($k\leq 2$) of $\phi$, 
then $\Theta (\phi,\delta\phi)$ is independent of
the choice of coordinates on $M$ and $E$
and thus the presymplectic form 
$\Omega_\Sigma(\phi, \delta_1\phi, \delta_2\phi)$ 
is then coordinate invariant. 
Moreover, 
note that $L(\phi)$ can be freely changed by 
addition of a locally constructed exact form $d\mu(\phi)$,
without affecting the field equations $\E(\phi)$. 
This changes $\Theta(\phi,\delta\phi)$ by 
addition of a locally constructed $(n-1)$-form $d\delta\mu(\phi)$,
but leaves 
$\Omega_\Sigma(\phi, \delta_1\phi, \delta_2\phi)$
unchanged. 
Therefore, up to its dependence on $\Sigma$,
the symplectic structure 
$\Omega_\Sigma(\phi, \delta_1\phi,\delta_2\phi)$ 
is uniquely determined by $\E(\phi)$ in this situation. 

Now consider a complete, nowhere vanishing vector field 
$\xi$ on $M$. 
It will be assumed that there exists 
a well-defined Lie derivative acting on $\phi$
associated to the diffeomorphism generated by $\xi$ on $M$.
Let $\Sigma$ be a connected region contained in 
a fixed hypersurface in $M$ 
with a closed boundary $\partial\Sigma$. 
(Note, if $\Sigma$ is simply-connected,
$\partial\Sigma$ is a closed $n-2$-surface in $M$ bounding $\Sigma$. 
If $\Sigma$ is multiply-connected, 
then $\partial\Sigma$ is a disjoint union of closed $n-2$-surfaces. 
Also, if $\Sigma$ extends to ``infinity'', 
then $\partial\Sigma$ contains a corresponding 
``asymptotic boundary'' $n-2$-surface.)

\Proclaim{ Definition A.1.}{
A {\it Hamiltonian} conjugate to $\xi$ on $\Sigma$ 
is a function $H_\Sigma(\xi;\phi)=\int_\Sigma \Hdens(\xi;\phi)$
for some locally constructed $(n-1)$-form $\Hdens(\xi;\phi)$
such that, on solutions $\Phi$, 
\EQ
H'_\Sigma(\xi;\Phi,\delta\phi) 
= \Omega_\Sigma(\Phi, \delta\phi,\Lie{\xi}\Phi) 
\endEQ
where $\Lie{\xi}$ denotes the Lie derivative,
and $H'_\Sigma(\xi;\phi,\delta\phi) \equiv \delta H_\Sigma(\xi;\phi)$. }

This is a covariant formulation of 
the standard Hamiltonian symplectic structure,
with the ``time'' direction defined by $\xi$,
called the {\it time flow} vector field. 
(In particular, \Ref{Wald-Lee} outlines a construction of
a standard phase space and Hamiltonian equations of motion
determined from this covariant structure.)
Note that a Hamiltonian, if it exists, 
is automatically conserved along $\xi$ for solutions $\Phi$,
\ie/ $\Lie{\xi} H_\Sigma(\xi;\Phi)=0$.
It will now be shown that 
when $\xi$ is a ``time symmetry'' of the Lagrangian $L(\phi)$,
then a Hamiltonian conjugate to $\xi$ exists 
and is simply given by the Noether charge associated to $\xi$. 

Given any vector field $\zeta$ on $M$, 
consider the variation 
$\delta_\zeta\phi\equiv\Lie{\zeta}\phi$. 
If this is a symmetry of the Lagrangian, 
so that 
\EQ\label{Lsymm}
\delta_\zeta L(\phi) = L'(\phi,\Lie{\zeta}\phi)
=d(i_\zeta L(\phi)) = \Lie{\zeta}L(\phi) 
\endEQ
by means of the identity $\Lie{\zeta} L(\phi) = d(i_\zeta L(\phi))$,
then one can define a conserved {\it Noether current} $(n-1)$-form 
$J(\zeta;\phi)$ by 
\EQ
J(\zeta;\phi)=\Theta(\phi,\Lie{\zeta}\phi)-i_\zeta L(\phi)
\endEQ
where $i_\zeta$ is the interior product. 
Conservation of this current simply means that, 
on solutions $\Phi$, 
$J(\zeta;\phi)$ is closed
\EQ
dJ(\zeta;\Phi) 
= d\Theta(\Phi,\Lie{\zeta}\Phi)-d(i_\zeta L(\Phi))
= L'(\Phi,\delta\Phi) -\Lie{\zeta} L(\Phi)
=0 
\endEQ
through the symmetry condition \eqref{Lsymm}. 
The integral of $J(\zeta;\Phi)$ over $\Sigma$ 
defines the {\it Noether charge}
\EQ
Q_\Sigma(\zeta)= \int_\Sigma J(\zeta;\Phi) . 
\endEQ
One finds that the ``time'' derivative of this charge 
with respect to $\xi$
is given by 
\EQ
\Lie{\xi} Q_\Sigma(\zeta)
=\int_\Sigma \Lie{\xi} J(\zeta;\Phi)
= \int_\Sigma i_\xi d J(\zeta;\Phi) +d(i_\xi J(\zeta;\Phi))
= \oint_{\partial\Sigma} i_\xi J(\zeta;\Phi)
\endEQ
where $i_\xi J(\zeta;\phi)$ is called the {\it flux} 
of the Noether current. 
Hence, if the flux vanishes on $\partial\Sigma$,
then the charge is conserved
for solutions $\Phi$. 

Examples of field theories 
which admit a symmetry $\delta_\zeta\phi=\Lie{\zeta}\phi$ 
are 
(i) any generally-covariant theory 
on a fixed, background spacetime $(M,g)$
with an isometry vector field $\zeta$
(\ie/ $\Lie{\zeta}g=0$),
where $L(\phi)$ is purely a function of 
$g$, $\phi$ and its metric-covariant derivatives $\nabla\phi$; 
(ii) any diffeomorphism-covariant theory,
whose field variables $\phi$ include the spacetime metric $g$,
where $L(\phi)$ is purely a function of 
$\phi$, curvature tensor of $g$, 
and their metric-covariant derivatives. 

For a diffeomorphism-covariant theory, 
$\delta_\zeta\phi$ is a symmetry for all vector fields $\zeta$. 
Consequently, since $J(\zeta;\phi)$ 
is locally constructed out of $\zeta$, 
one can show that in this case 
\cite{Wald-Lee}
\EQ
J(\zeta;\Phi)=dQ(\zeta;\Phi)
\endEQ
for some locally constructed $(n-2)$-form $Q(\zeta;\phi)$, 
called the {\it Noether current potential}. 
Then the Noether charge reduces to a surface integral
\EQ
Q_\Sigma(\zeta)= \int_\Sigma J(\zeta;\phi)
= \oint_{\partial\Sigma} Q(\zeta;\phi) . 
\endEQ

In contrast, for a generally-covariant theory, 
$J(\zeta;\Phi)$ is related to 
the conserved stress-energy tensor $T(\phi)$
defined by considering variations of $g$, 
\EQ
*\delta_g L(\phi) = -\frac{1}{2} T(\phi) \delta g +*d\Theta(\phi,\delta g) . 
\endEQ
One can show that \cite{Wald-Iyer1}, 
on solutions $\Phi$,
\EQ
J(\zeta;\Phi) = *i_\zeta T(\Phi) +d \tau(\zeta;\Phi)
\endEQ
for some locally constructed $(n-2)$-form $\tau(\zeta;\phi)$. 

\Proclaim{ Proposition A.2. }{
For any symmetry $\delta_\zeta\phi=\Lie{\zeta}\phi$
admitted by a Lagrangian $L(\phi)$, 
the field equations and symplectic potential 
satisfy
\EQs
&& 
\delta_\zeta \E(\phi) = \Lie{\zeta} \E(\phi) , 
\label{covLieE}\\
&& 
\delta_\zeta \Theta(\phi,\delta\phi) 
=  \Lie{\zeta}\Theta(\phi,\delta\phi) +d \psi(\zeta;\phi,\delta\phi) 
\label{covvarT}
\endEQs
where $\psi(\zeta;\phi,\delta\phi)$ 
is some locally constructed $(n-2)$-form. }

\proclaim{ Proof :}
Consider an arbitrary variation of 
the Lagrangian symmetry condition \eqref{Lsymm}, 
\EQ\label{varsymmL}
0=\delta( \delta_\zeta L(\phi) - \Lie{\zeta}L(\phi) )
= \delta_\zeta\delta L(\phi) - \Lie{\zeta}\delta L(\phi) . 
\endEQ
From \Eqref{varL}, one has
\EQ
\Lie{\zeta}( \delta L(\phi) )
= \Lie{\zeta}(\E(\phi) \delta\phi)  
+ \Lie{\zeta} d \Theta(\phi,\delta\phi)
= (\Lie{\zeta}\E(\phi)) \delta\phi 
+ \E(\phi) \delta\Lie{\zeta}\phi
+  d\Lie{\zeta} \Theta(\phi,\delta\phi) , 
\endEQ
and similarly
\EQ
\delta_\zeta( \delta L(\phi) )
= (\delta_\zeta\E(\phi)) \delta\phi 
+ \E(\phi) \delta\Lie{\zeta}\phi
+  d\delta_\zeta \Theta(\phi,\delta\phi) , 
\endEQ
since $\delta_\zeta\phi=\Lie{\zeta}\phi$. 
Hence, \Eqref{varsymmL} yields
\EQ\label{varsymmLeq}
( \delta_\zeta\E(\phi) - \Lie{\zeta}\E(\phi) ) \delta\phi 
= d( \Lie{\zeta} \Theta(\phi,\delta\phi)
- \delta_\zeta \Theta(\phi,\delta\phi) )
\endEQ
holding for all $\delta\phi$. 
By taking $\delta\phi$ to have compact support
and integrating the equation \eqref{varsymmLeq} over $M$,
one obtains 
$\int_M ( \delta_\zeta\E(\phi) - \Lie{\zeta}\E(\phi) ) 
\delta\phi =0$
which immediately yields \Eqref{covLieE}. 
Then \Eqref{varsymmLeq} shows that 
$\Lie{\zeta} \Theta(\phi,\delta\phi)
- \delta_\zeta \Theta(\phi,\delta\phi)$
is a closed $(n-1)$-form holding for all $\phi$. 
Since this expression is locally constructed in terms of $\phi$, 
it follows that \cite{Wald,Anderson}
\Eqref{covvarT} holds. 
\endproof

From these results,
one finds that 
the variation of the Noether current is given by 
\EQs
J'(\zeta;\phi,\delta\phi) \equiv \delta J(\zeta;\phi)
&& 
= \delta\Theta(\phi,\Lie{\zeta}\phi) - i_\zeta \delta L(\phi)
\nonumber\\&&
= \omega(\phi,\delta\phi, \Lie{\zeta}\phi) 
+\delta_\zeta\Theta(\phi,\delta\phi)
- i_\zeta( d\Theta(\phi,\delta\phi) + \E(\phi)\delta\phi )
\nonumber\\&&
= \omega(\phi,\delta\phi, \Lie{\zeta}\phi) 
-i_\zeta( \E(\phi)\delta\phi )
+ d ( i_\zeta \Theta (\phi,\delta\phi) +\psi(\zeta;\phi,\delta\phi) )
\label{varJid}
\endEQs
using the identity
$i_\zeta( d\Theta(\phi,\delta\phi) )
= \Lie{\zeta}\Theta(\phi,\delta\phi) 
- d ( i_\zeta \Theta (\phi,\delta\phi) )$. 

\Proclaim{ Lemma A.3. }{
On solutions $\Phi$,
\EQ
\Omega_\Sigma(\Phi,\Lie{\zeta}\Phi, \delta\phi)
= -\int_\Sigma J'(\zeta;\Phi,\delta\phi) 
+\oint_{\partial\Sigma}  
i_\zeta \Theta (\Phi,\delta\phi)
+\psi(\zeta;\Phi,\delta\phi) . 
\endEQ
Thus, for variations $\delta\phi$
with compact support in the interior of $\Sigma$,
\ie/ $\delta\phi|_{\partial\Sigma}=0$,
\EQ
\Omega_\Sigma(\Phi,\Lie{\zeta}\Phi, \delta\phi)
= -\int_\Sigma J'(\zeta;\Phi,\delta\phi) . 
\endEQ }

One can then apply this result to the time flow vector field $\zeta=\xi$
to obtain a Hamiltonian. 

\Proclaim{ Theorem A.4. }{
The Noether current $J(\xi;\phi)$ yields 
a Hamiltonian conjugate to $\xi$ on $\Sigma$
given by $H_\Sigma(\xi;\phi)=\int_\Sigma J(\xi;\phi)$
under compact support variations $\delta\phi$. 
For solutions $\Phi$,
$H_\Sigma(\xi;\Phi)=Q_\Sigma(\xi)$
is the conserved Noether charge
associated to $\xi$. }

For variations $\delta\phi$ without compact support,
there exists a Hamiltonian if and only if 
one can find a locally constructed $(n-2)$-form $B(\xi;\phi)$ 
such that
\EQ\label{defB}
\oint_{\partial\Sigma} 
B'(\xi;\Phi,\delta\phi) 
-i_\xi\Theta (\Phi,\delta\phi) 
-\psi(\xi;\Phi,\delta\phi)
=0
\endEQ
where $B'(\xi;\phi,\delta\phi) \equiv \delta B(\xi;\phi)$. 
If one restricts to variations $\delta\phi=\delta\Phi$, 
then by considering a second variation and antisymmetrizing
in this equation,  
one obtains the necessary condition 
\EQ
\oint_{\partial\Sigma}
\delta_1( i_\xi\Theta (\Phi,\delta_2\Phi) 
+ \psi(\xi;\Phi,\delta_2\Phi) )
- \delta_2( i_\xi\Theta (\Phi,\delta_1\Phi) 
+ \psi(\xi;\Phi,\delta_1\Phi) )
=0
\endEQ
for existence of $B(\xi;\phi)$. 
This condition can also be shown to be sufficient
\cite{Wald-Zoupas}. 

\Proclaim{ Definition A.5. }{
An allowed boundary condition on $\phi$ 
is a set of field components $\bc(\phi)|_{\partial\Sigma}$ 
locally constructed from $\phi$, partial derivatives $\partial^k\phi$,
and spacetime quantities associated to $\xi,\Sigma,\partial\Sigma$, 
such that for all variations $\delta\phi$ 
satisfying $\bc'(\phi,\delta\phi)|_{\partial\Sigma}=0$,
where $\bc'(\phi,\delta\phi)\equiv \delta\bc(\phi)$,
there exists a Hamiltonian $H_\Sigma(\xi;\phi)$
conjugate to $\xi$ on $\Sigma$. }

One now has the following main result. 

\Proclaim{ Theorem A.6. }{ 
A Hamiltonian conjugate to $\xi$ on $\Sigma$ 
exists under variations $\delta\phi$ 
without compact support 
if and only if 
\EQ
\oint_{\partial\Sigma} i_\xi \omega(\Phi, \delta_1\Phi,\delta_2\Phi) 
=\oint_{\partial\Sigma} \psi'(\xi;\Phi, \delta_1\Phi,\delta_2\Phi)
\endEQ
on solutions $\Phi$,
where
$\psi'(\xi;\phi, \delta_1\phi,\delta_2\phi) \equiv 
\delta_1\psi(\xi;\phi,\delta_2\phi)
- \delta_2\psi(\xi;\phi,\delta_1\phi)$. 
This determines the allowed boundary conditions 
$\bc(\phi)|_{\partial\Sigma}$
for the field equations to admit 
a covariant Hamiltonian formulation. 
Then the Hamiltonian is
\EQ\label{noetherH}
H_\Sigma(\xi;\phi)=
\int_\Sigma J(\xi;\phi) - d B(\xi;\phi)
\endEQ
with $B(\xi;\phi)$ given by \Eqref{defB}
up to an arbitrary function of the boundary data $\bc(\phi)$ and $\xi$.
Furthermore, under the allowed boundary conditions, 
the Hamiltonian and symplectic structure
are independent of choice of $\Sigma$. }

The surface integral 
$\oint_{\partial\Sigma} i_\xi \omega(\Phi, \delta_1\Phi,\delta_2\Phi)$
will be referred to as the {\it symplectic flux} through $\partial\Sigma$.

For a diffeomorphism-covariant theory, 
or a generally-covariant theory on a background spacetime,
one can show that $\psi(\xi;\phi,\delta\phi)\equiv 0$. 
Hence the necessary and sufficient condition
for existence of a Hamiltonian becomes
\EQ
\oint_{\partial\Sigma} i_\xi \omega(\Phi, \delta_1\Phi,\delta_2\Phi) 
=0
\endEQ
and, furthermore, from the relation \eqref{defB} between 
$\Theta (\Phi,\delta\phi)$ and $B(\xi;\phi)$, 
it follows that one has
\EQ
B(\xi;\phi) |_{\partial\Sigma} 
= ( i_\xi {\tilde B}(\phi) )|_{\partial\Sigma}
\endEQ
where ${\tilde B}(\phi)$ is a locally constructed $(n-1)$-form. 
Then on solutions $\Phi$ 
the Hamiltonian takes the following form:
in the case of a diffeomorphism-covariant theory, 
\EQ
H_B(\xi;\Phi)=
\oint_{\partial\Sigma} Q(\xi;\Phi) - i_\xi {\tilde B}(\Phi)
\endEQ
which is a surface integral;
and in the case of a generally-covariant theory,
\EQ
H(\xi;\Phi)+H_B(\xi;\Phi)= 
 \int_\Sigma *i_\zeta T(\Phi)
+ \oint_{\partial\Sigma} \tau(\xi;\Phi) -i_\xi {\tilde B}(\Phi)
\endEQ
where $H(\xi;\Phi)= \int_\Sigma *i_\zeta T(\Phi)$ 
is the canonical energy associated to $\Phi$ on $\Sigma$,
and $H_B(\xi;\Phi)$ is the surface integral term. 

To conclude, some further features of the Noether charge Hamiltonian
will now be developed.

\Proclaim{ Definition A.7.}{
The coefficient of an arbitrary compact support variation
$\delta\phi|_\Sigma$ in the equation
\EQ\label{varH}
\int_\Sigma 
\Omega_\Sigma(\phi, \delta\phi,\Lie{\xi}\phi) 
- H'_\Sigma(\xi;\phi,\delta\phi) 
\equiv \int_\Sigma \HE(\xi;\phi) \delta\phi 
=0 
\endEQ
yields the Hamiltonian field equations for $\phi$, 
$\HE(\xi;\phi)=0$. }

\Proclaim{ Theorem A.8.}{
The Hamiltonian field equations $\HE(\xi;\phi)=0$
are equivalent to the Lagrangian field equations $\E(\phi)=0$. }

\proclaim{ Proof :}
The Hamiltonian \eqref{noetherH} satisfies the variational identity
\EQ\label{varHid}
\Omega_\Sigma(\phi, \delta\phi,\Lie{\xi}\phi) 
- H'_\Sigma(\xi;\phi,\delta\phi) 
= \int_\Sigma i_\xi( \E(\phi)\delta\phi )
\endEQ
derived from \Eqrefs{varJid}{defB}.
Hence, for arbitrary compact support variations $\delta\phi|_\Sigma$,
$\E(\phi)=0$ holds if and only if $\phi$ satisfies $\HE(\xi;\phi)=0$.
\endproof

A field variation, denoted by $\delta\phi\nondyn$,
is a {\it symplectic degeneracy direction}
if $\Omega_\Sigma(\phi,\delta\phi,\delta\phi\nondyn)=0$
holds for arbitrary compact support variations $\delta\phi$. 
Such degeneracies arise whenever the Lagrangian $L(\phi)$ 
admits a gauge symmetry
(\ie/ a symmetry $\delta_\chi \phi$ that is locally constructed from
$\phi$, partial derivatives $\partial^k\phi$, 
and that depends linearly on a set of parameters $\chi$
freely specifiable as functions on $M$.)
Note that the set $\{\delta\phi\nondyn\}$ of all degeneracy directions
is a vector space.
Then, a nondegeneracy direction, denoted by $\delta\phi\dyn$,
is represented as an equivalence class in the vector space of 
all field variations $\{\delta\phi\}$ quotiented by 
all symplectic degeneracy directions $\{\delta\phi\nondyn\}$, 
namely $\delta\phi\dyn=\delta\phi/\delta\phi\nondyn$.
This decomposition yields a break up of 
the Hamiltonian field equations \eqref{varH}
into non-dynamical constraint equations,
\EQ\label{constraintHeq}
H'_\Sigma(\xi;\phi,\delta\phi\nondyn)=0 , 
\endEQ
and dynamical evolution equations, 
\EQ\label{dynamicalHeq}
H'_\Sigma(\xi;\phi,\delta\phi\dyn)=
\Omega_\Sigma(\phi, \delta\phi\dyn,\Lie{\xi}\phi) ,
\endEQ
through arbitrary variations 
$\delta\phi\nondyn,\delta\phi\dyn$ 
with compact support on $\Sigma$.

Since it assumed that the set $\{\phi\}$ of all fields 
has a linear (vector bundle) structure, 
the symplectic degeneracy directions $\delta\phi\nondyn$ 
can be identified with a corresponding set of field components,
denoted $\phi\nondyn$,
which will be called non-dynamical with respect to $\Sigma$.
Similarly, the nondegeneracy directions $\delta\phi\dyn$
determine a set of equivalence classes of field components,
denoted $\phi\dyn$,
which will be called dynamical with respect to $\Sigma$.
(Note these components $\phi\dyn$ and $\phi\nondyn$ are 
locally constructed from $\phi$, partial derivatives $\partial^k\phi$,
and spacetime quantities associated to $\Sigma$.)
Then, from the Hamiltonian variational identity \eqref{varHid},
one can view the constraint equations \eqref{constraintHeq}
and evolution equations \eqref{dynamicalHeq}
as arising equivalently through 
the action principle \eqref{action}
by variations with respect to $\phi\nondyn$ and $\phi\dyn$. 

In summary, the Noether charge formalism presented here 
gives a covariant Hamiltonian formulation 
for Lagrangian field theories 
in the situation where the underlying time flow 
is given by a symmetry of the Lagrangian.


\begin{references}
\def\v#1{{\bf #1}}

\bibitem{Regge-Teitelboim}
T. Regge and C. Teitelboim,
``Role of surface integrals in the Hamiltonian Formulation of
General Relativity'',
Annals of Physics \v{88} (1974) 286-318.

\bibitem{ADM}
R. Arnowitt, S. Deser, C.W. Misner,
``The Dynamics of General Relativity'',
in {\it Gravitation: An Introduction to Current Research}
ed. L. Witten (New York: Wiley) (1962).

\bibitem{Witten}
C. Crnkovi\'c and E. Witten,
in {\it Three Hundred Years of Gravitation},
ed. S.W. Hawking and W. Israel
(Cambridge University Press: Cambridge) (1987) p.676-684. 

\bibitem{Wald-Lee}
J. Lee and R.M. Wald,
``Local symmetries and constraints'',
J. Math. Phys. \v{31}, (1990) 725-743.

\bibitem{GIMSY}    
M. Gotay, J. Isenberg, and J. E. Marsden,  ``Momentum Maps and Classical
Relativistic Fields'', Part I (1997);
Part II (1999), Unpublished Notes.    

\bibitem{Nester}  
J. M. Nester, ``A Covariant Hamiltonian for Gravity Theories'',
Mod. Phys. Lett. A \v{6}, (1991) 2655-2661;
J. M. Nester, ``Some Progress in Canonical Gravity'', in Directions in
General Relativity,
ed. B. L. Hu, M. P. Ryan and C. V. Vishveshwara (Cambridge University
Press: Cambridge) Vol 1 (1993) 245-260.

\bibitem{Kijowski}
J. Kijowski, ``A Simple Derivation of Canonical Structure and Quasi-Local
Hamiltonians in General Relativity'',
Gen. Relativ. Gravit. \v{29}, (1997) 307-343;  
J. Kijowski and W.M. Tulczyjew,
{\it A Symplectic Framework for Field Theories},
Lecture Notes in Physics No. 107
(Springer-Verlag: Berlin) (1979).

\bibitem{symplecticvectors}
S.C. Anco and R.S. Tung,
``Properties of the symplectic structure of General Relativity
for spatially bounded spacetime regions'',
J. Math. Phys. \v{43}, (2002) 3984-4019.

\bibitem{Wald-Iyer1}
V. Iyer and R.M. Wald,
``Some Properties of Noether Charge and a Proposal for Dynamical
Black Hole Entropy'',
Phys. Rev. D \v{50}, (1994) 846-864. 

\bibitem{Wald-Zoupas}
R.M. Wald and A. Zoupas,
``A General Definition of ``Conserved Quantities'' in General Relativity
and Other Theories of Gravity'',
Phys. Rev. D \v{61}, (2000) 084027. 

\bibitem{Wald-Iyer2}
V. Iyer and R.M. Wald,
``A comparison of Noether charge and Euclidean methods for Computing
the Entropy of Stationary Black Holes'',
Phys. Rev. D \v{52}, (1995) 4430-4439. 

\bibitem{Wald-book}
R.M. Wald,
{\it General Relativity}
(University of Chicago Press: Chicago) (1984).

\bibitem{Jackson}
J.D. Jackson,
{\it Classical Electrodynamics}
(Wiley: New York) 2nd edition (1975).

\bibitem{Wald2}
R.M. Wald, 
``The first law of black hole mechanics'', 
in {\it Directions in General Relativity} Volume 1, 
ed. B.L. Hu, M. Ryan, C.V. Vishveshwara, 
(Cambridge University Press: Cambridge), (1993). 

\bibitem{York}
J.W. York, 
``Boundary terms in the Action Principles of General Relativity'',
Found. Phys. \v{16}, (1986) 249-257.

\bibitem{Brown-York1}
J.D. Brown and J.W. York,
``Quasilocal energy and conserved charges derived from the 
gravitational action'', 
Phys. Rev. D \v{47}, (1993) 1407-1419. 

\bibitem{Brown-York2}
J.D. Brown, S.R. Lau, J.W. York,
``Action and Energy of the Gravitational Field'',
Annals Phys. \v{297} (2002) 175-218.

\bibitem{Nester1}
C.M. Chen, J.M. Nester, R.S. Tung, 
``Quasilocal energy-momentum for geometric gravity theories'',
Phys. Lett. A \v{203} (1995), 5-11.

\bibitem{Nester2}
C.M. Chen and J.M. Nester, 
``Quasilocal quantities for GR and other theories of gravity'',
Class. Quant. Grav. \v{16} (1999), 1279-1304. 

\bibitem{Friedrich}
H. Friedrich and G. Nagy,
``The Initial Boundary Value Problem for Einstein's Vacuum Field
Equation'',
Comm. Math. Phys. \v{201} (1999) 619-655.

\bibitem{Wald}
R.M. Wald,
J. Math. Phys. \v{31} (1990) 2378-2384. 

\bibitem{Anderson}
I.M. Anderson,
``Introduction to the variational bicomplex''
in {\it Mathematical Aspects of Classical Field Theory},
ed. M. Gotay, J. Marsden, V. Moncrief,
Cont. Math. \v{132} (1992) 51-73.








\end{references}
\end{document}